\documentclass[10pt,
							 twocolumn,
							 superscriptaddress,
							 english,
							 pra,
							 showpacs,
							 floatfix,
							 aps
							]{revtex4-1}
\usepackage[utf8]{inputenc}
\usepackage{amsmath}
\usepackage{amssymb}
\usepackage{graphicx}
\usepackage{xspace}
\usepackage{mathtools}

\makeatletter
\@ifundefined{textcolor}{}
{%
 \definecolor{BLACK}{gray}{0}
 \definecolor{WHITE}{gray}{1}
 \definecolor{RED}{rgb}{1,0,0}
 \definecolor{GREEN}{rgb}{0,1,0}
 \definecolor{BLUE}{rgb}{0,0,1}
 \definecolor{CYAN}{cmyk}{1,0,0,0}
 \definecolor{MAGENTA}{cmyk}{0,1,0,0}
 \definecolor{YELLOW}{cmyk}{0,0,1,0}
}

\usepackage[normalem]{ulem}

\usepackage{soul}
\usepackage{braket}
\usepackage[backref=none,
bookmarksnumbered=true,
bookmarks=true,
bookmarksopen=true,
colorlinks=true,
citecolor=blue,
linkcolor=blue,
anchorcolor=green,
urlcolor=blue,unicode=false]{hyperref}

\let\baraccent=\= 
\renewcommand{\=}[1]{\stackrel{#1}{=}} 

\makeatother

\usepackage{babel}

\begin{document}
\title{Yu-Shiba-Rusinov states in real metals}

\author{Felix von Oppen}
\affiliation{\mbox{Dahlem Center for Complex Quantum Systems and Fachbereich Physik, Freie Universit\"at Berlin, 14195 Berlin, Germany}}

\author{Katharina J. Franke}
\affiliation{\mbox{Fachbereich Physik, Freie Universit\"at Berlin, 14195 Berlin, Germany}}
\date{\today}
\begin{abstract}
Theoretical descriptions of Yu-Shiba-Rusinov (YSR) states induced by magnetic impurities inside the gap of a superconductor typically rely on a classical spin model or are restricted to spin-$\frac{1}{2}$ quantum spins. These models fail to account for important aspects of YSR states induced by transition-metal impurities, including the effects of higher quantum spins coupled to several conduction-electron channels, crystal or ligand-field effects, and magnetic anisotropy. We introduce and explore a zero-bandwidth model, which incorporates these aspects, is readily solved numerically, and analytically tractable in several limiting cases. The principal simplification of the model is to neglect Kondo renormalizations of the exchange couplings between impurity spin and conduction electrons. Nevertheless, we find excellent correspondence in those cases, in which we can compare our results to existing numerical-renormalization-group calculations. We apply the model to obtain and understand phase diagrams as a function of pairing strength and magnetic anisotropy as well as subgap excitation spectra. The single-channel case is most relevant for transition-metal impurities embedded into metallic coordination complexes on superconducting substrates, while the multi-channel case models transition-metal adatoms.
\end{abstract}
\pacs{%
			} 
\maketitle 


\section{Introduction}

While magnetic impurities in superconductors are a long-standing topic in condensed matter physics, there has been a recent resurgence of interest, fueled in part by possible sightings of topological superconductivity in chains of magnetic adatoms on superconducting substrates \cite{NadjPerge2014, Ruby2015chains, Pawlak2016, Feldman2017, Jeon2017,Kim2019}. Scanning tunneling microscopy (STM) addresses individual magnetic adatoms on superconductors and directly resolves the Yu-Shiba-Rusinov (YSR) states, which the adatoms induce in the superconducting substrate at subgap energies \cite{Yazdani1997, Heinrich2018}. Experimental advances such as the use of superconducting STM tips have greatly improved the resolution \cite{Ji2008, Franke2011} and experiments now routinely identify multiple subgap YSR states \cite{Ruby2016, Hatter2015, Choi2017,Liebhaber2019}, probe the hybridization of YSR states in adatom dimers \cite{Ruby2018, Choi2018, Kezilebieke2018,Beck2020}, or explore larger assemblies of adatoms, for instance in the search of Majorana bound states of one-dimensional \cite{Ruby2017, Kamlapure2018, Schneider2020} or the Majorana edge modes of two-dimensional topological superconductors \cite{Menard2017,Palacio2019, Menard2019}. 

In many of these experiments, the magnetic impurities of choice are transition metals, either directly as adatoms \cite{Yazdani1997,Ji2008,Ruby2016,Choi2018} or integrated as metal centers into coordination complexes \cite{Franke2011,Hatter2015,Hatter2017,Kezilebieke2018,Farinacci2018, Malavolti2018, Brand2018,Etzkorn2018}. These systems are typically characterized by higher spins (up to $S_{i}=5/2$ for $d$ electron systems), crystal or ligand-field splittings, and single-ion anisotropy, either due to the substrate or the coordination complex. While these effects are of crucial importance in experiment, they are frequently sidestepped by minimal models in theoretical works on YSR states. Crystal fields have been shown to induce multiple YSR states in higher-spin systems \cite{Ruby2016, Choi2017} and magnetic anisotropy has been explored by numerical renormalization group calculations for a limited set of parameters \cite{Zitko2011}. While these numerical renormalization group calculations give important guidance, much of the current understanding remains descriptive due to the lack of analytical approaches. 

It is the purpose of the present paper to develop a framework for understanding YSR states of transition-metal impurities in real metals within a minimal model, clearly revealing the underlying physics and allowing extensions to many cases of experimental interest. The purpose of the model is not to quantitatively predict detailed properties for specific impurities, but rather to provide qualitative insights into phase diagrams and subgap excitation spectra of higher-spin impurities when crystal or ligand fields and anisotropies are present. Even though the model neglects Kondo renormalizations, it qualitatively reproduces and explains many observations made in much heavier numerical renormalization group calculations.

\begin{figure*}[]
 	\includegraphics[width=0.72\textwidth]{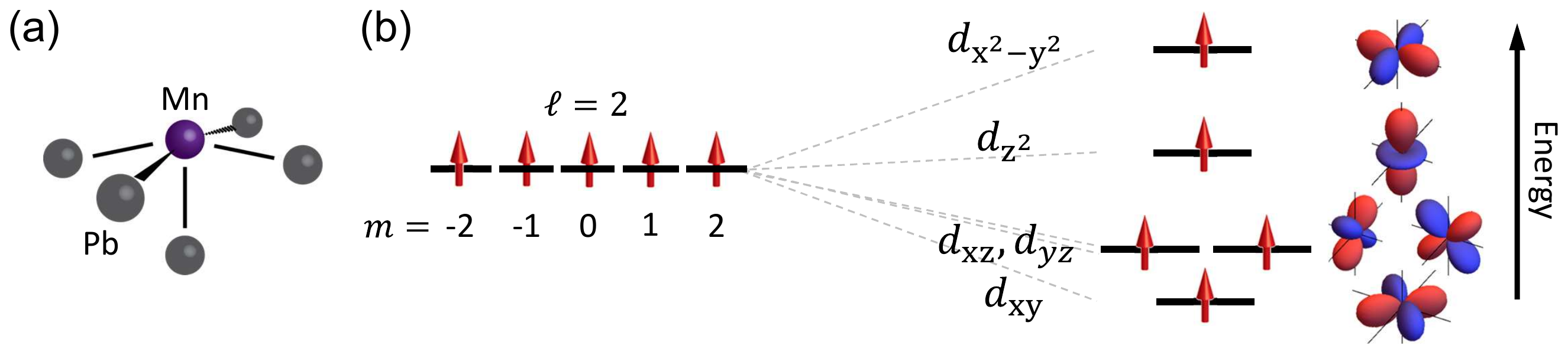}\hspace{.3cm}
 	\includegraphics[width=0.22\textwidth]{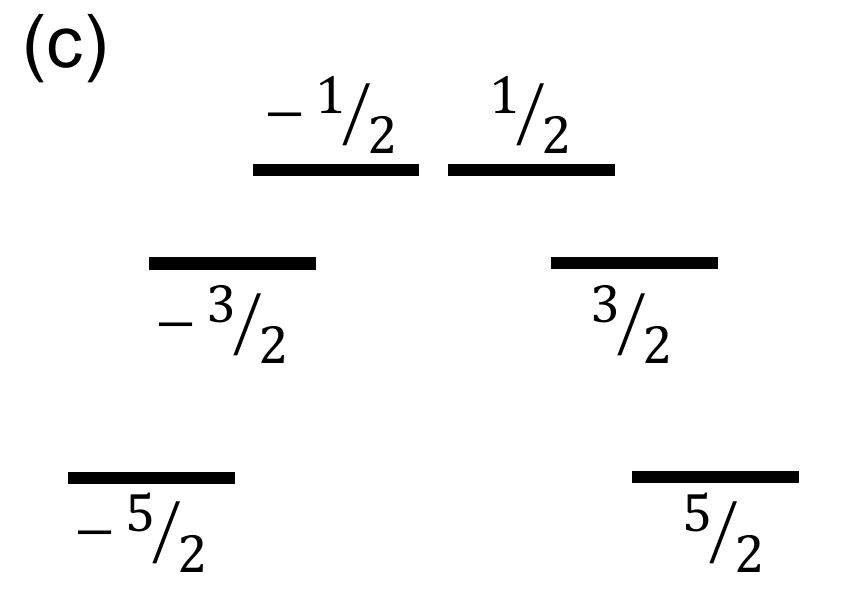}
	\caption{Illustration of crystal-field effects for spin-$\frac{5}{2}$ Mn adatoms on a Pb(001) substrate \cite{Ruby2016}. (a) Adsorption site of Mn on Pb(001) has $C_{4v}$ point-group symmetry. (b) Crystal-field splitting of the $\ell=2$ manifold. According to Hund's rules for the $d^5$ configuration of Mn, all $d$ orbitals are singly occupied, which  results in an $S_i=\frac{5}{2}$ spin. Each ligand-field-split $d$ orbital couples to a separate conduction-electron channel. The point-group symmetry implies that the $d_{xz}$ and $d_{yz}$ orbitals are degenerate and the exchange couplings with their respective conduction-electron channels are identical. (c) Bare anisotropy splitting of an $S_i=\frac{5}{2}$ spin for uniaxial anisotropy $D<0$ (easy-axis anisotropy). The levels are labeled by the projections $M$ of the impurity spin. (For easy-plane anisotropy, $D>0$, the order of the levels would be reversed.) The effect of the anisotropy is strongly modified, when the impurity binds quasiparticles on a superconducting substrate.} 
	\label{fig_LigandField}
\end{figure*}

As emphasized by Schrieffer \cite{Schrieffer1967}, transition-metal impurities do not exchange scatter conduction electrons in the $\ell=0$ angular-momentum channel, as assumed in most models of YSR states, but rather in the $\ell=2$ channel. This implies that  the impurity spin couples to up to five (rather than one) conduction-electron channels. The detailed nature of the symmetry-adapted conduction-electron channels as well as the crystal-field splitting of the impurity $d$ orbitals are governed by the point group at the impurity location. For a quenched orbital moment and strong Hund coupling, the impurity exchange scatters the conduction electrons within each channel coupled to a singly-occupied $d$ orbital, so that there are $2S_{i}$ channels for an impurity spin of magnitude $S_{i}$ (see Fig.\ \ref{fig_LigandField} for an illustrative example based on Ref.\ \cite{Ruby2016}). Taking this picture as a starting point \cite{Moca2008,Ruby2016}, one generally expects the appearance of multiple YSR states, with the number of YSR states and their degeneracies reflecting the symmetries of the crystal field. For adatoms, there is strong experimental support for the validity of this picture from a direct observation of the YSR wave functions in STM maps, which closely resemble appropriate $d$ orbitals \cite{Ruby2016, Beck2020}. 

As a further consequence of crystal fields, YSR states are also subject to single-ion anisotropy \cite{Zitko2011}. On normal-metal substrates, magnetic anisotropy regularly leads to the appearance of spin excitations in tunneling spectra, resulting from transitions between different eigenstates of the anisotropy Hamiltonian of the impurity spin \cite{Hirjibehedin2007,Tsukuhara2009}. According to these experiments, typical anisotropy energies can be comparable to superconducting energy gaps, making magnetic anisotropy a relevant perturbation of standard models of YSR states. Numerical renormalization group calculations show that the ground state and the subgap excitation spectra of magnetic adatoms on superconductors are indeed sensitive to the type and magnitude of the magnetic anisotropy \cite{Zitko2011}. This is further supported by experiments on a transition-metal complex for which the observation of multiple YSR resonances has been interpreted in terms of single-ion-anisotropy-induced splittings \cite{Hatter2015, Hatter2017}. 

Experiments on magnetic adatoms and molecules show that adatom spins are typically quantum mechanical. This follows most directly from the observation of Kondo resonances, both on normal-metal substrates \cite{Madhavan1998,Li1998} and on superconductors  \cite{Franke2011,Hatter2017,Kamlapure2018,Farinacci2020, Verdu2021, Kamlapure2019,Odobesko2020}, as well as the discrete nature of spin excitations \cite{Hirjibehedin2007,Tsukuhara2009, Heinrich2013b, Kezilebieke2019}. At the same time, YSR states are frequently described within models which treat the adatom spin as classical \cite{Yu1965,Shiba1968,Rusinov1969}. While these models are quite successful \cite{Flatte1997PRL,Flatte1997,Salkola1997,Balatsky2006,Pientka2013,Poyhonen2014,
Hoffman2015,Kirsanskas2015}, they fall short in a number of ways. For instance, they fail to predict the correct degeneracies of the many-body ground state or incorrectly suggest that YSR binding energies should be independent of the sign of the exchange interaction between adatom spin and substrate electrons. Important insight into the YSR states of quantum impurities comes from approximate analytical calculations \cite{Soda1967,Zittarz1970,Kirsanskas2015} as well as numerical studies \cite{Satori1992,Bauer2007,Hecht2008,Zitko2011,Yao2014}. With few exceptions \cite{Zitko2011}, these studies of quantum spins have focused on spin-$\frac{1}{2}$ impurities, precluding considerations of crystal-field or magnetic-anisotropy effects. 

Assemblies of magnetic adatoms on superconductors have been studied as platforms for realizing topological superconductivity. Most prominently, signatures of Majorana bound states have been observed in chains of Fe adatoms on a Pb substrate \cite{NadjPerge2014, Ruby2015chains, Pawlak2016, Feldman2017, Jeon2017}. In these experiments, the adatoms are closely spaced, and it is believed that the emergence of topological superconductivity in these structures is possible due to the direct hybridization of the Fe $d$ orbitals \cite{Li2014a,Peng2015}. It would also be extremely interesting to realize topological superconductivity in more dilute assemblies of adatoms, with the adatoms sufficiently spaced out so that the direct overlap of their $d$ orbitals can be neglected \cite{Pientka2013}. In this case, each adatom induces YSR states in the superconducting substrate which would then hybridize with one another. Clearly, a detailed understanding of YSR states in real metals would be highly desirably for designing and interpreting corresponding experiments, which may then open the path to designer topological superconductors by STM manipulation of adatoms. 

\section{Model}

\subsection{Exchange coupling}

The impurity spin $\mathbf{S}_{i}$ emerges from the interplay of Hund coupling and crystal- or ligand-field splittings. High-spin configurations are favored when the Hund coupling exceeds the crystal field splitting (or the combined ligand- and cystal-field splittings in metal coordination complexes), while low-spin configurations result in the opposite case. For atomic impurities, the crystal-field splitting of the $d$ orbitals is controlled by the point group of the impurity site. The split sets of $d$ levels and their degeneracies can be understood in terms of irreducible representations of the point group (Fig.\ \ref{fig_LigandField}). Similarly, it is convenient to work with conduction-band states, which transform according to irreducible representations of the point group. Hybridization of impurity and conduction-band states will then conserve the representation label $m$. For bulk impurities, it is natural to expand the conduction-band states in spherical waves and only $\ell=2$ partial waves exchange scatter from the impurity \cite{Schrieffer1967}. Assuming symmetry-adapted $d$ orbitals and conduction-electron channels for adatoms, the antiferromagnetic exchange coupling takes the form
\begin{equation}
    H_\mathrm{ex}=  \sum_m J_m \mathbf{S}_{i}\cdot \psi_{m}^\dagger(\mathbf{0})\mathbf{s} \psi_{m}^{\phantom{\dagger}}(\mathbf{0}),
\end{equation}
where $m$ labels the channels. If two channels are symmetry related, their exchange couplings $J_m$ are identical. The expression for the exchange coupling is valid when orbital angular momentum is quenched, so that there is only spin exchange. Here, $\psi_{m}(\mathbf{0})$ denotes the spinor of conduction electron field operators for partial waves in representation $m$, evaluated at the impurity position, and $\mathbf{s}=\frac{1}{2}\boldsymbol{\sigma}$ is the electron spin in terms of the vector of Pauli matrices $\boldsymbol{\sigma}$ in spin space. The sum over $m$ extends over the half-filled $d$-electron orbitals of the adatom, so that the impurity spin is exchange coupled to $2S_{i}$ independent conduction-electron channels, with a different exchange coupling $J_m>0$ for each set of symmetry-related $d$ orbitals. 

For impurity spins associated with adsorbed transition-metal complexes, the molecular ligand field may not have the same symmetry as the crystal field imposed by the substrate. Moreover, the metal center will be typically lifted slightly off the surface by the molecular scaffold. In this situation, hybridization to conduction band electrons may be dominated by the $d_{z^2}$ orbital, as this orbital extends furthest towards the substrate, and originates predominantly from Bloch wave vectors $\mathbf{k}$ directed perpendicular to the surface, whose evanescent tail extends furthest out of the substrate. The reduction in symmetry will tend to induce a complete splitting between channels and a single channel may have a much larger exchange coupling than the others.  

\subsection{YSR states of classical impurities}
\label{sec:classical}

The Hamiltonian of the impurity spin coupled to the substrate superconductor takes the form
\begin{eqnarray}
    H &=& \int d\mathbf{r} \sum_m \psi_m^\dagger(\mathbf{r})\left[ (\epsilon_\mathbf{p} -\mu ) \tau_z + \Delta \tau_x \right.
\nonumber\\
    && \,\,\,\,\,\,\,\,\,\,\,\,\,\,\,\,\,\,\,\,\,\,
  +\left.  (V_m\tau_z + J_m\mathbf{S}_{i}\cdot\mathbf{s}) \delta(\mathbf{r}) \right]\psi_m(\mathbf{r}).
\end{eqnarray}
Here, we included the amplitude $V_m$ for potential scattering in all channels $m$, while $\epsilon_\mathbf{p}$ denotes the normal-state dispersion of the substrate electrons, $\mu$ is the chemical potential, $\Delta$ the superconducting pairing, and $\boldsymbol{\tau}$ the vector of Pauli matrices in particle-hole space. For all channels, the electron field operator $\psi_m$ is a Nambu spinor $\psi = [\psi_\uparrow^{\phantom{\dagger}}, \psi_\downarrow^{\phantom{\dagger}},\psi^\dagger_\downarrow,-\psi_\uparrow^\dagger]^T$.

YSR states are most commonly described within a model, which describes the magnetic impurity spin $\mathbf{S}_{i}$ as classical \cite{Yu1965,Shiba1968,Rusinov1969}. In this model, the different conduction-electron channels decouple and the impurity induces a single pair of YSR states in each channel with energies $\pm \epsilon_m$, where
\begin{equation}
   \epsilon_m = \Delta \frac{1-\alpha_m^2+\beta_m^2}{\sqrt{(1-\alpha_m^2+\beta_m^2)^2+4\alpha_m^2}}
\end{equation}
in terms of the dimensionless exchange coupling $\alpha_m=\pi \nu_0 JS_{i}/2$ ($\nu_0$ is the normal-state density of states at the Fermi energy) and the dimensionless strength $\beta=\pi\nu_0 V$ of potential scattering. Potential scattering leads to a difference between the orbital electron and hole wave functions of the YSR state. 

As a function of the exchange coupling, the YSR states traverse the gap and cross at zero energy for a critical value of $\alpha_m^2=1+\beta_m^2$. This zero-energy crossing indicates a quantum phase transition, at which the fermion parity of the ground state changes \cite{Sakurai1970}. Beyond this critical value, the impurity binds a quasiparticle, which (partially) screens the impurity spin. A classical spin breaks the time reversal symmetry of the full quantum model, and the many-body ground states are nondegenerate on both sides of the quantum phase transition. 

\subsection{Few-site model for quantum impurities}

\subsubsection{Spin-$\frac{1}{2}$ impurity}

When the impurity spin is classical, we can choose a coordinate system such that $\mathbf{S}_i=S_{i,z}\mathbf{\hat z}$ and the exchange coupling is purely longitudinal $\sim {S}_{i,z}{s}_z$. This reduction to a longitudinal exchange coupling does not carry over to quantum spins, for which the transverse exchange $\sim\frac{1}{2}[{S}_{i,+}{s}_-+{S}_{i,-}{s}_+]$ plays an important role. For instance, either the unscreened or the screened state of the quantum model must be Kramers degenerate, so that the quantum phase transition is generically associated with a change in the ground-state degeneracy. This follows because time reversal symmetry is unbroken in the full quantum model and the system has half-integer total spin on one side of the transition.

This is correctly captured by a zero-bandwidth model for an $S_i=\frac{1}{2}$ quantum impurity, in which the superconductor has only a single site (see, e.g., Ref.\ \cite{Kirsanskas2015,Affleck2000,Vecino2003,Oguri2004,Tanaka2007,Grove2018,Bouman2020,Estrada2020} for applications to Josephson junctions and quantum dot systems),
\begin{equation}
   H = \Delta (c^\dagger_\downarrow       
       c^\dagger_\uparrow +  c^{\phantom{\dagger}}_\uparrow c^{\phantom{\dagger}}_\downarrow) + \sum_{\sigma\sigma'} c_\sigma^\dagger  [V \tau_z\delta_{\sigma\sigma'} + J \mathbf{S}_{i}
      \cdot   \mathbf{s}^{\phantom{\dagger}}_{\sigma\sigma'}] c^{\phantom{\dagger}}_{\sigma'}.
\end{equation}
Here, $c_\sigma$ denotes the annihilation operator of an electron with spin $\sigma$ on the superconducting site and $V$ models the potential scattering amplitude associated with the impurity. This single-site Hamiltonian is time reversal symmetric, preserves fermion parity, and has full spin-rotation symmetry. Eigenstates can thus be classified according to magnitude and $z$-component of the total spin
\begin{equation}
   \mathbf{S}=\mathbf{S}_{i} + \sum_{\sigma\sigma'} c_\sigma^\dagger   \mathbf{s}^{\phantom{\dagger}}_{\sigma\sigma'} c^{\phantom{\dagger}}_{\sigma'}
\end{equation}
as well as fermion parity. 

The even-parity subspace is spanned by states in which the site is empty -- states $|S_{i}=\frac{1}{2}, M = \pm\frac{1}{2} \rangle \otimes |0\rangle$ -- or doubly occupied  -- states $|\frac{1}{2}, M \rangle \otimes |2\rangle$ with $|2\rangle = c^\dagger_\uparrow c^\dagger_\downarrow|0\rangle$. None of these states has a net electronic spin and the exchange coupling has only zero matrix elements within this subspace. Thus, the impurity spin remains unscreened and free, and there is a Kramers doublet of low-energy states $|\frac{1}{2}, M\rangle\otimes|\mathrm{BCS}\rangle$, where the electronic state is the BCS state 
\begin{equation}
  |\mathrm{BCS}\rangle = u|0\rangle + v |2\rangle
\end{equation}
with electron and hole amplitudes $u^2=\frac{1}{2}[1+V/\sqrt{V^2+\Delta^2}]$ and $v^2=\frac{1}{2}[1-V/\sqrt{V^2+\Delta^2}]$, respectively. The (many-body) energy of these low-energy states is $E_e=V - \sqrt{V^2+\Delta^2}$. We note that the second set of degenerate eigenstates $|\frac{1}{2}, M\rangle\otimes (v|0\rangle -u |2\rangle)$ are  above-gap excited states with two Bogoliubov quasiparticles, which will play no role in the following. 

In the odd-parity subspace spanned by $|\frac{1}{2}, M\rangle \otimes |\frac{1}{2}, \sigma\rangle$ with $|\frac{1}{2}, \sigma\rangle = c_\sigma^\dagger |0\rangle$, pairing is ineffective. (Note that we write the state of the impurity spin first, followed by the electronic state, for which we find it convenient to indicate the magnitude of the spin in addition to the spin projection $\sigma=\pm\frac{1}{2}$.) Due to spin-rotation symmetry, the impurity and electron spins couple into singlet and triplet states, with the singlet state having the lower energy. Thus, there is only a single low-energy state
\begin{equation}
    |\mathrm{sing}\rangle = \frac{1}{\sqrt{2}}(|\frac{1}{2},\frac{1}{2}\rangle \otimes |\frac{1}{2},-\frac{1}{2}\rangle - |\frac{1}{2},-\frac{1}{2}\rangle \otimes |\frac{1}{2}, \frac{1}{2}\rangle)
\end{equation} 
with energy $E_o=V-3J/4$. The energy of the triplet state equals $E_o=V+J/4$ and corresponds to an above-gap state, which we do not consider. 

The even- and odd-fermion-parity states become degenerate, when $3J/4=\sqrt{V^2+\Delta^2}$. This defines the quantum phase transition between a Kramers doublet of even-fermion-parity ground states with free and unscreened impurity spin, and a singlet ground state with screened impurity spin. For weak exchange coupling, $3J/4<\sqrt{V^2+\Delta^2}$, the even-fermion-parity BCS states form the doublet ground state, and the odd-fermion-parity singlet state the subgap YSR excitation with excitation energy $\epsilon=\sqrt{V^2+\Delta^2}-3J/4$. Conversely, the singlet odd-fermion-parity state becomes the ground state for strong exchange coupling, $3J/4 > \sqrt{V^2+\Delta^2}$, and the YSR excitation is the even-fermion-parity doublet with excitation energy $\epsilon=3J/4-\sqrt{V^2+\Delta^2}$.

\subsubsection{Spin-$S_i$ impurity}

In the remainder of this paper, we will discuss a generalized version of this zero-bandwidth model for higher-spin impurities subject to crystal or ligand fields and single-ion anisotropy. In this model, each of the $2S_{i}$ conduction-electron channels of the superconductor is represented by a separate single-site superconductor labeled by $m$. Thus, the Hamiltonian of the model takes the form
\begin{eqnarray}
   H &=&\sum_m \Delta (c^\dagger_{m\downarrow}       
       c^\dagger_{m\uparrow} +  c^{\phantom{\dagger}}_{m\uparrow} c^{\phantom{\dagger}}_{m\downarrow}) 
   \nonumber\\
 &&+            \sum_{\sigma\sigma'}\sum_m c_{m\sigma}^\dagger    [V_m\tau_z\delta_{\sigma\sigma'} + J_m \mathbf{S}_{i}
      \cdot    \mathbf{s}^{\phantom{\dagger}}_{\sigma\sigma'}] c^{\phantom{\dagger}}_{m\sigma'}
  \nonumber\\
   && +D S_{{i},z}^2 +E (S_{{i},x}^2-S_{{i},y}^2) .
\label{model_hamiltonian}
\end{eqnarray}
Here, the last two terms describe the single-ion anisotropy, with $D$ and $E$ quantifiying the axial and transverse anisotropies, respectively. There are many cases, in which the transverse anisotropy is symmetry forbidden. For this reason, we retain only the axial anisotropy $D$ throughout most of this paper, and comment on the effects of $E$ only occasionally, when a nonvanishing $E$ introduces qualitative changes. The spectrum of the anisotropy term by itself is illustrated in Fig.\ \ref{fig_LigandField}c for the case of easy-axis anisotropy, $D<0$, which favors large projections $M$ of the impurity spin. Easy-plane anisotropy, $D>0$, in contrast, favors small impurity-spin projections $M$. 

In many ways, the quantum model goes over into the classical model in the limit of large easy-axis anisotropy. In this limit, the anisotropy essentially freezes the impurity spin into the $M=\pm S_i$ states. Then, the exchange coupling is dominated by the longitudinal contribution and the transverse contribution to the exchange coupling becomes suppressed. However, even in this limit the binding of a quasiparticle remains associated with a change in the ground-state degeneracy of the quantum model. This is a consequence of the existence of the two large-spin-projection states $M=\pm S_i$, and leads to characteristic differences in the excitation spectra of the classical and quantum models as we will see below. It should also be kept in mind that even in this case, the classical-spin model does not capture Kondo renormalizations of the exchange coupling strength due to processes above the energy scale of the anisotropy energy.

It is the main point of this paper to show that the model in Eq.\ (\ref{model_hamiltonian}) is remarkably useful to treat the YSR physics of higher-spin quantum impurities coupled to several channels and subject to single-ion anisotropy. The model can be viewed as the limit of the full problem, in which $\Delta$ becomes large compared to the bandwidth, but it is important to keep its limitations in mind. We close this section by commenting on aspects that are not captured. To start with, the model clearly cannot properly describe above-gap excitations. Due to the single-site nature of the superconducting channels, there is no quasiparticle continuum of the superconductor, with which above-gap states of the impurity could hybridize. We will thus systematically ignore such above-gap states and focus exclusively on subgap excitations. Even for subgap states close to the superconducting gap edge, the zero-bandwidth approximation does not fully capture level-repulsion effects.

The absence of a quasiparticle continuum also implies that the model neglects Kondo renormalizations of the exchange coupling. Thus, the exchange couplings entering into the model should be interpreted as renormalized couplings. In superconductors, scaling to stronger coupling (for antiferromagnetic exchange) is ultimately cut off at the superconducting gap. Moreover, the renormalized exchange coupling can depend on the strength of the magnetic anisotropy, which lifts the degeneracy of the $2S_{i}+1$ impurity-spin states (see Sec.\ \ref{sec:beyond} for further discussion of this point). 

Finally, restricting the superconductor to a single site eliminates the spatial wave-function structure of the YSR states, which can be probed directly in experiment \cite{Ruby2016}. 

\section{YSR states of metal coordination complexes: Single-channel model}

We begin by considering higher-spin magnetic impurities coupled to a single channel [i.e., we consider $S_{i}\geq 1$, setting $J_1=J$ and all other $J_m=0$ in Eq.\ (\ref{model_hamiltonian})]. This situation has been treated by extensive numerical renormalization group calculations \cite{Zitko2011}, which allows us to probe the validity of the zero-bandwidth model. Moreover, as argued above, this situation will frequently constitute a useful model for discussing YSR states of transition-metal complexes on superconducting substrates due to the dominant exchange coupling in one channel. 

For a single channel, the model Hamiltonian in Eq.\ (\ref{model_hamiltonian}) reduces to 
\begin{eqnarray}
   H &=&\Delta (c^\dagger_{\downarrow}       
       c^\dagger_{\uparrow} +  c^{\phantom{\dagger}}_{\uparrow} c^{\phantom{\dagger}}_{\downarrow}) 
   +            \sum_{\sigma\sigma'} J \mathbf{S}_{i}
      \cdot c_{\sigma}^\dagger   \mathbf{s}^{\phantom{\dagger}}_{\sigma\sigma'} c^{\phantom{\dagger}}_{\sigma'}
  \nonumber\\
   && +D S_{{i},z}^2 +E (S_{{i},x}^2-S_{{i},y}^2) .
\label{single_channel_hamiltonian}
\end{eqnarray}
Here, we set $V=0$ for all channels, neglecting potential scattering by the impurity for simplicity. Compared to the spin-$\frac{1}{2}$ model already discussed, the new aspect is that magnetic anisotropy becomes relevant and we can discuss phase diagrams as well as  excitation spectra as a function of magnetic anisotropy. 

To classify the eigenstates as well as the phases of the model, it is useful to consider the symmetries of the model. The general model in Eq.\ (\ref{model_hamiltonian}) conserves the fermion parity of each channel separately, while spin rotation symmetry is completely broken by the anisotropy. However, the model remains invariant under spin rotations about the $\hat z$ axis for vanishing transverse anisotropy $E$. In this case, the $z$-component of the total spin
\begin{equation}
   \mathbf{S}=\mathbf{S}_{i} + \sum_{\sigma\sigma'} \sum_m c_{m\sigma}^\dagger   \mathbf{s}^{\phantom{\dagger}}_{\sigma\sigma'} c^{\phantom{\dagger}}_{m\sigma'}
\end{equation}
remains a good quantum number. The ground states can then be labeled by $(Q,S_z)$, where $S_z$ is the projection of total spin and 
\begin{equation}
  Q = \frac{1}{2}\sum_m [1 - (-1)^{n_m}] 
\end{equation}
with $n_m=\sum_\sigma c^\dagger_{m\sigma}c_{m\sigma}$ denotes the number of quasiparticles bound to the impurity. (Notice that in the ground state, the quasiparticles will always be bound in those channels, which have the largest exchange couplings. Thus, $Q$ specifies the fermion parities $(-1)^{n_m}$ of all channels.) In the single-channel model in Eq.\ (\ref{single_channel_hamiltonian}), $Q$ takes on the values $0$ and $1$, distinguishing between even- and odd electron-number eigenstates. 

\begin{figure*}[]
	\includegraphics[width=0.98\textwidth]{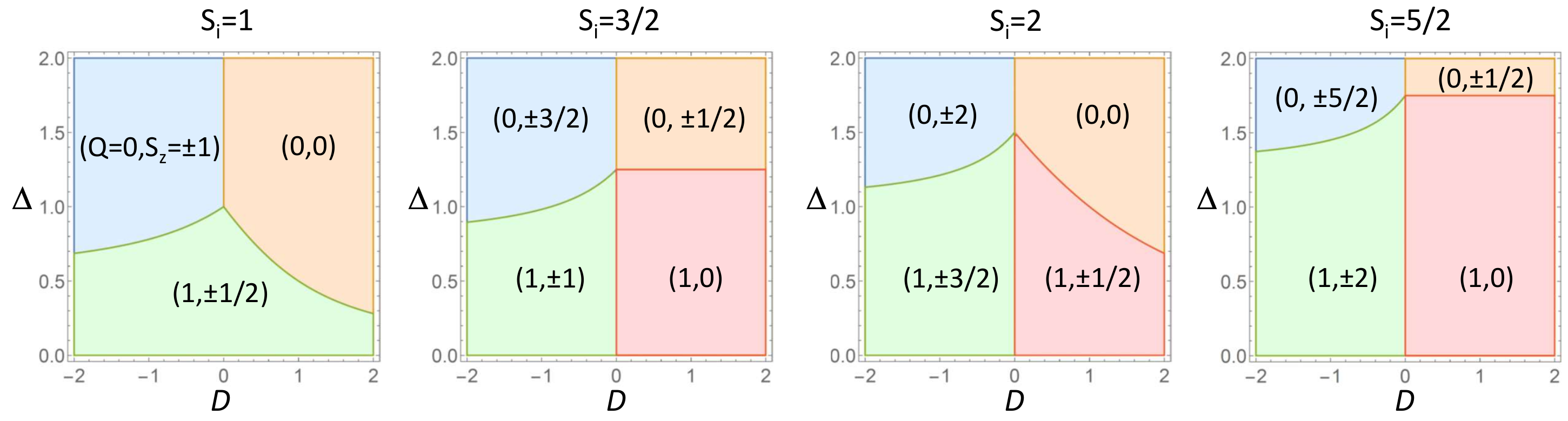}
	\caption{Phase diagrams for various impurity spins $S_{i}$ as a function of pairing strength $\Delta$ and uniaxial anisotropy $D$, based on the single-channel model in Eq.\ (\ref{single_channel_hamiltonian}). Transverse anisotropy is set to zero and the exchange coupling $J=1$ sets the energy scale. Phases are labeled by $(Q,S_z)$, where $Q$ is the number of bound quasiparticles (fermion parity) and $S_z$ the projection of the total spin. The phases at larger pairing strengths $\Delta$ have an unscreened impurity spin and no bound quasiparticle ($Q=0$, even fermion parity). The phases at lower pairing strengths $\Delta$ have a partially screened impurity spin and a bound quasiparticle ($Q=1$, odd fermion parity). See the appendix for corresponding spectra of the Hamiltonian in Eq.\ (\ref{single_channel_hamiltonian}).}
	\label{fig:phase_diagram_single_channel}
\end{figure*}

\subsection{Phase diagrams}
\label{sec:mol_phase_diagrams}

Figure \ref{fig:phase_diagram_single_channel} shows phase diagrams as a function of pairing strength $\Delta$ and uniaxial anisotropy $D$ for various impurity spins,
obtained by numerical diagonalization of the single-channel model in Eq.\ (\ref{single_channel_hamiltonian}). Since $E=0$, the phases are labeled by $(Q,S_z)$. Larger pairing energies favor $Q=0$ states, corresponding to unscreened, even-fermion-parity ground states. The spin projection of these states depends on the sign of the anisotropy $D$. For $D<0$, the anisotropy favors large spin projections, and the unscreened ground state has $S_z=\pm S_{i}$. For $D>0$, the anisotropy favors small spin projections, and the unscreened ground state has  $S_z=0$ for integer $S_{i}$ or $S_z=\pm \frac{1}{2}$ for half-integer $S_{i}$.

For smaller pairing energies $\Delta$, the impurity binds a quasiparticle, resulting in partially screened, odd-fermion-parity ground states with $Q=1$. The quasiparticle is bound to the impurity by the exchange interaction, which grows with the impurity spin $S_i$ for our definitions. Consequently, the partially screened phases become more prominent with increasing impurity spin $S_z$. (The exchange coupling $J$ is fixed to unity for all panels.) Just as for the unscreened phases, the spin projections of the partially screened ground states depend on the sign of the uniaxial anisotropy. For $D<0$, the spin projections equal $\pm(S_i-\frac{1}{2})$ and are reduced in magnitude compared to the unscreened phase. This is a consequence of the fact that the exchange interaction binding the quasiparticle to the impurity is antiferromagnetic. For $D>0$, the binding of the quasiparticle implies that the ground state has $S_z=0$ for half-integer $S_{i}$ and $S_z=\pm \frac{1}{2}$ for integer $S_{i}$, just reversed compared to the unscreened phases. The case of $S_i=1$ is exceptional in that there is only a single partially screened phase, independent of the sign of $D$. The reason is that in this case, the partially screened phase has $S_z=\pm\frac{1}{2}$, so that the anisotropy causes no energy splitting.

The effects of anisotropy differ qualitatively for positive and negative uniaxial anisotropies. Negative anisotropy, $D<0$, systematically favors the unscreened phase. The phase boundary moves to smaller pairing energies as $D<0$ increases in magnitude,  tending towards a nonzero value of $\Delta$ for $D\to-\infty$. This trend of the phase boundary results from the increasing polarization of the impurity spin along the $\hat z$-direction as $D$ becomes more negative. This suppresses the contribution of the transverse part of the exchange interaction, $J(S_{i,x}s_x+S_{i,y}s_y)$, reducing the gain in exchange energy of the quasiparticle.

Positive anisotropies, $D>0$, consistently favor, or at least stabilize, the phases with zero spin projection. For integer impurity spins, it is the unscreened phase which has zero spin projection, and the phase boundary to the partially screened phase moves to smaller pairing energies $\Delta$ as $D$ increases. The phase boundary ultimately approaches $\Delta=0$ and the partially screened phase disappears completely as $D\to\infty$. This suppression of the screened phase is absent for half-integer impurity spins. In this case, the phase boundary between the unscreened and partially screened phases is insensitive to the anisotropy, and the partially screened phase with zero spin projection is robust. This striking difference in behavior can be traced to the Kramers degeneracy, which necessarily exists for half-integer spins. Integer impurity spins have a unique ground state for large and positive anisotropy, and the exchange coupling affects the energy only in second order in perturbation theory. The resulting binding energy is thus proportional to $J^2/D$ and vanishes as $D\to\infty$. In contrast, the ground state is Kramers degenerate for half-integer spins and the exchange interaction acts in first order in perturbation theory, resulting in binding energies of order $J$. 

There are various quantum phase transitions between the ground states in Fig.\ \ref{fig:phase_diagram_single_channel}. Fermion parity changes beween the unscreened and the partially screened phases. The existence of these quantum phase transitions is robust when modifying the Hamiltonian. In contrast, fermion parity remains unchanged at the quantum phase transitions at $D=0$ between states with different spin projections. First, it should be noted that right at these fermion-parity-preserving quantum phase transitions, additional spin projections can be degenerate with the two phases (for the unscreened phases, for instance,  $S_z=\pm 1$ for $S_i=2$ and $S_z=\pm \frac{1}{2},\pm\frac{3}{2}$ for $S_i=5/2$). Second, these quantum phase transitions are absent if the Hamiltonian fully breaks spin rotation symmetry, e.g., due to a nonzero transverse anisotropy $E$. In this case, the two phases can no longer be distinguished as the spin projection $S_z$ ceases to be a good quantum number. 

\begin{figure*}[]
	\includegraphics[width=0.85\textwidth]{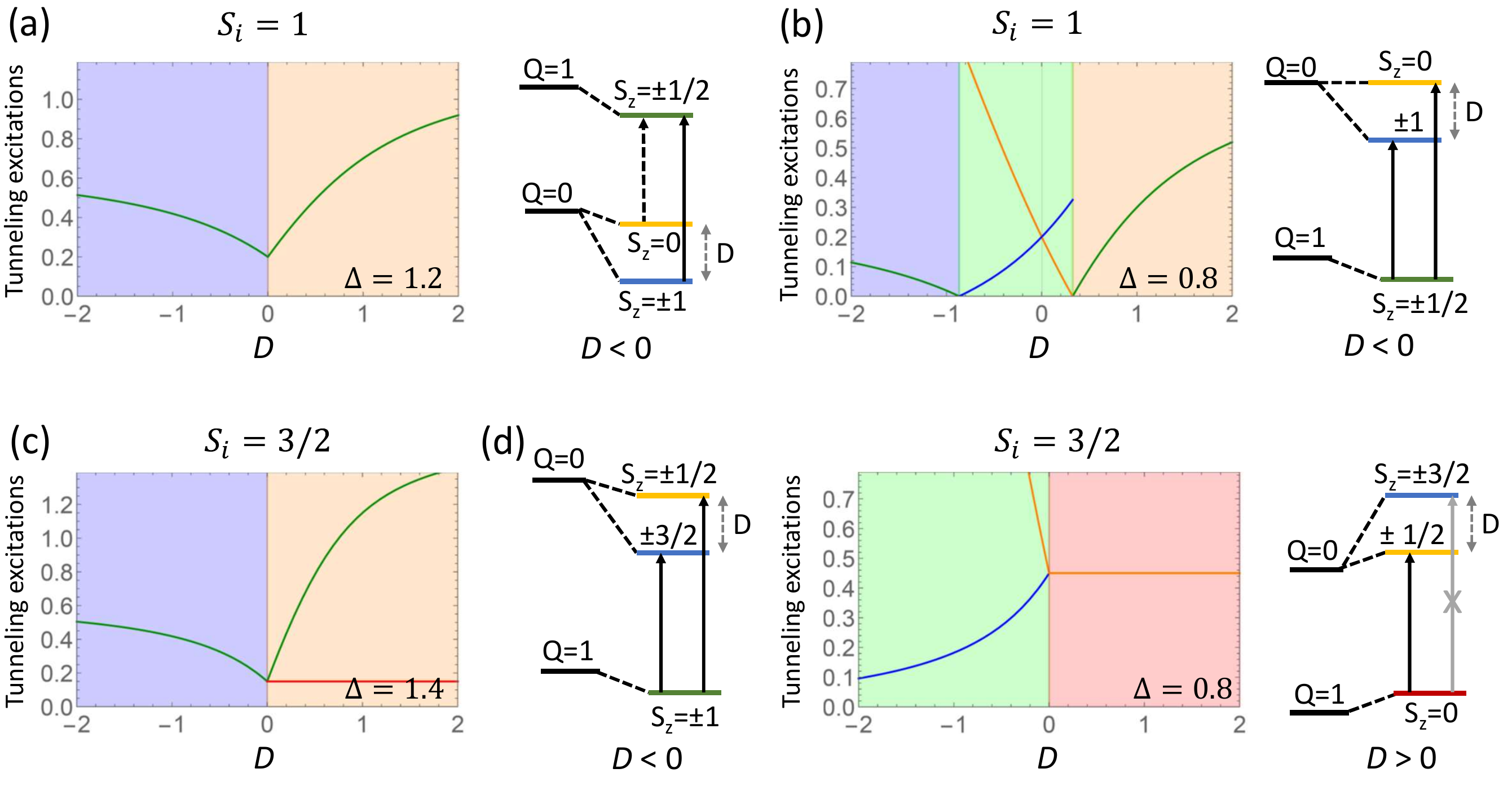}
	\caption{Subgap tunneling excitation energies of the single-channel model in Eq.\ (\ref{single_channel_hamiltonian}) as a function of anisotropy $D$  for $S_{i}=1$ (panels a and b) and for $S_{i}=3/2$ (panels c and d). Background color indicates the ground state and line color the excited state (same color coding as in the corresponding phase diagram in Fig.\ \ref{fig:phase_diagram_single_channel}). Transverse anisotropy is set to zero and the exchange coupling $J=1$ sets the energy scale. Pairing strength $\Delta$ is chosen as indicated in the panels. The maximal excitation energies shown correspond to the gap edge $\Delta$ of the superconductor. (a) Tunneling excitations out of the unscreened ground states for $S_i=1$. The corresponding term diagram for $D<0$ uses the same color scheme. Allowed transitions out of the ground state, shown as full arrows, are included in the excitation spectrum. The anisotropy splitting generates additonal excitations out of excited states, shown as a dashed arrow. These appear at finite temperatures, with their intensity showing an activated temperature dependence. (b) Tunneling excitations for $S_i=1$ at smaller pairing energy, including excitations out of the screened ground state (green background). The corresponding term diagram for $D<0$ shows that for a screened ground state, the YSR excitations exhibit an anisotropy splitting even at zero temperature. (c) Tunneling excitations out of the unscreened ground states for $S_i=3/2$. (d) Tunneling excitations out of the screened ground states for $S_i=3/2$. Anisotropy splitting is seen at negative anisotropy $D<0$, see left term diagram. For positive anisotropy $D>0$, the anisotropy splitting cannot be observed due to the selection rule $\Delta S_z=\pm \frac{1}{2}$, see right term diagram. In all tunneling spectra, cusps and terminations of excited-state energies are associated with quantum phase transitions. 
}
	\label{fig:excitations_single_channel}
\end{figure*}

\subsection{Excitation spectra}

Excitations can be investigated by fermion-parity-preserving probes such as microwaves. More commonly, one measures subgap tunneling spectra, which probe excitations from even- to odd-fermion-parity states, or vice versa. Subgap excitations appear as pairs of peaks at symmetric positive and negative bias voltages. For normal-metal tips, these peaks occur directly at $eV = \pm \epsilon$, where $\epsilon$ is the excitation energy of a YSR state. For superconducting tips, the peaks appear at voltages, which are offset by the tip gap, $eV = \pm (\epsilon + \Delta_\mathrm{tip})$. Below, we restrict attention to the excitation energies $\epsilon$, which appear universally in STM experiments, regardless of whether the YSR states are probed by normal-metal or superconducting tips, whether tunneling is dominated by single-electron or two-electron (Andreev) tunneling \cite{Ruby2015}, and whether the adsorbate is an adatom or a transition metal complex. 

In principle, one can also compute peak heights within the model. However, the physics of the peak heights is much less universal and sensitive to the nature of the potential scattering, of the tip, of the tunneling process, and of the adsorbate \cite{Ruby2015,Farinacci2020}. For instance, tunneling is dominated by single-electron processes for weak tip-substrate coupling and the peak heights at positive and negative bias voltages reflect single-particle spectral weights. For stronger tip-substrate coupling, tunneling proceeds by Andreev processes. The corresponding peak heights are no longer described by single-particle spectral weights and moreover depend on the nature of the tip \cite{Ruby2015}. In experiment, the peak heights also exhibit pronounced spatial variations, reflecting the YSR wave functions \cite{Ruby2016}. Clearly, single-site models are incapable of capturing these spatial dependencies. In view of its nonuniversal physics, we refrain from further discussion of the peak heights. 

\subsubsection{Tunneling spectra}

Figure \ref{fig:excitations_single_channel} illustrates the subgap tunneling excitation energies as a function of anisotropy $D$ for $S_i=1$ and $S_i=3/2$. Representative spectra for $S_i=1$ are shown in panel a for excitations out of the unscreened ground states and panel b for excitations out of the partially screened ground state. Anisotropy splits the $Q=0$ manifold (no bound quasiparticle) into $S_z=0$ and degenerate $S_z=\pm 1$ states. (The latter would be further split by a nonzero transverse anisotropy $E$.) The $Q=1$ manifold consists of two Kramers degenerate $S_z=\pm \frac{1}{2}$ states. For the unscreened ground states (Fig.\ \ref{fig:excitations_single_channel}a), the subgap tunneling spectrum shows a single YSR excitation into the $Q=1$ manifold, originating from the $S_z=\pm 1$ ($S_z=0$) state for negative (positive) $D$. Additional excitations associated with the anisotropy splitting of the $Q=0$ manifold would appear at finite temperatures (thermal peaks), where the excited states of the $Q=0$ manifold have a finite occupation probability, see the dashed arrow in the term diagram in Fig.\ \ref{fig:excitations_single_channel}a. 

The anisotropy splitting shows up directly in zero-temperature tunneling spectra, when exciting the system out of the screened ground state. Figure \ref{fig:excitations_single_channel}b shows the subgap tunneling spectrum at a lower pairing strength $\Delta$, for which the ground state is unscreened at large positive and negative anisotropies $D$ (blue and orange backgrounds), but partially screened at intermediate anisotropies (green background). We again find a single YSR excitation into a partially screened state in the regions with unscreened ground states, consistent with the discussion of panel a. However, there are two YSR excitations, when the ground state is partially screened. In this case, the excited $Q=0$ manifold is unscreened and split by the anisotropy. Tunneling excitations connect the partially screened $Q=1$ ground state with both of these anisotropy-split levels, resulting in two YSR excitations unless the splitting becomes so large that one of the excitations moves out of the subgap region. The corresponding excitations are also shown in the term dia\-gram included in Fig.\ \ref{fig:excitations_single_channel}b. 

\begin{figure*}[]
	\includegraphics[width=0.95\textwidth]{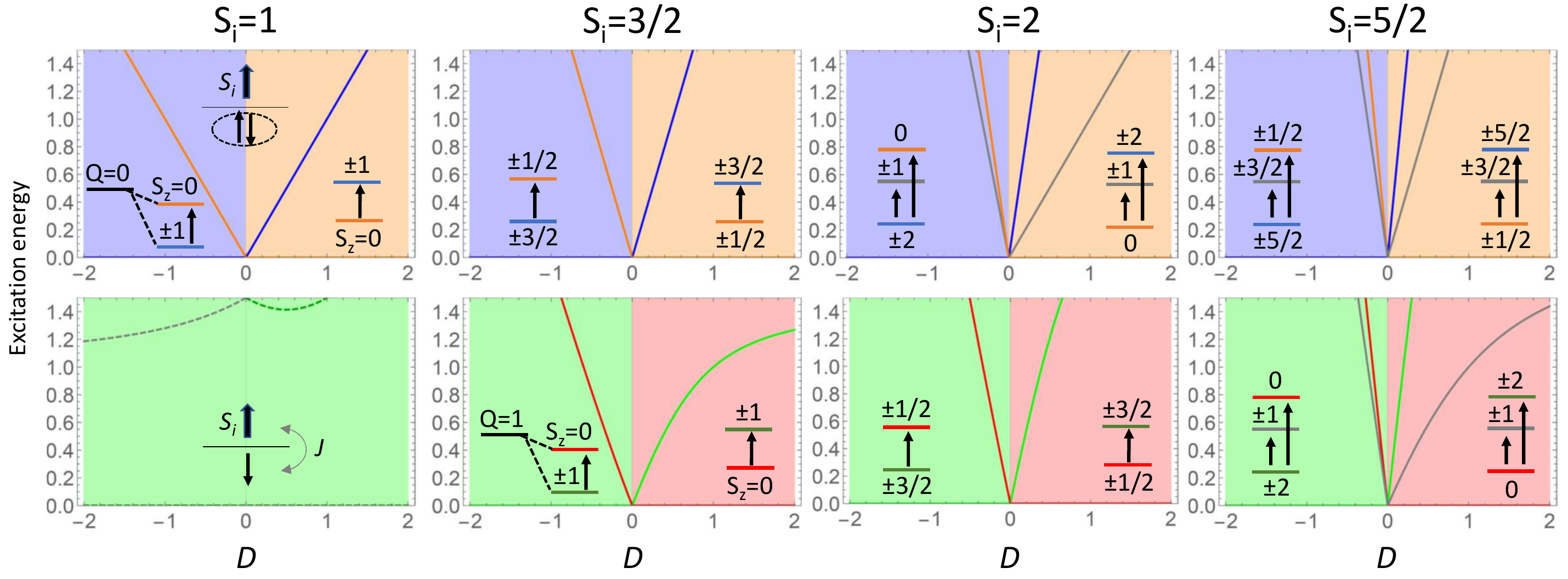}
	\caption{Subgap excitation spectra of the single-channel model in Eq.\ (\ref{single_channel_hamiltonian}) for fixed fermion parity as a function of anisotropy $D$  for impurity spins $S_{i}$ as indicated above the panels. Top row refers to even fermion parity states without quasiparticle ($Q=0$; illustrated by top inset in panel for $S_i=1$)
and the bottom row to odd fermion parity states with quasiparticle ($Q=1$; illustrated by bottom inset in panel for $S_i=1$). Background color indicates the ground state and line color the excited state (same color coding as in the corresponding phase diagram in Fig.\ \ref{fig:phase_diagram_single_channel}). Gray lines correspond to intermediate total spin projections, which do not become the ground state at any $D$. Dashed lines indicate  above-gap excitations (see Sec.\ \ref{sec:ppe}). The term diagrams illustrate the underlying transitions out of the ground states. In experiment, some transitions may be forbidden due to additional selection rules beyond $\Delta Q=0$, characteristic of the specific probe, and transitions out of excited states may appear at finite temperatures. Transverse anisotropy is set to zero and the exchange coupling $J=1$ sets the energy scale. The excitation spectra are independent of pairing strength $\Delta$, except that $\Delta$, here chosen as $\Delta=1.5$, determines the range, over which the states appear as subgap excitations. 
}
\label{fig:ParityPreservingExcitations_SingleChannel}
\end{figure*}

Quantum phase transitions can be associated with cusps in the excited-state spectrum, both at finite and at zero excitation energy (Figs.\ \ref{fig:excitations_single_channel}a and b, respectively), or even with terminations of excitations. Cusps occur at zero excitation energy, when the quantum phase transition is between states with $\Delta Q=\pm 1$. In this case, the ground-state and excited-state manifolds simply trade roles at  the quantum phase transition and one of the excitations necessarily reaches zero right at the transition. Cusps appear at finite excitation energies, when the quantum phase transition is between ground states with $\Delta Q\neq \pm 1$ and the YSR excitation excites the same manifold from both ground states. This situation occurs in Fig.\ \ref{fig:excitations_single_channel}a, where the system is excited into the partially screened manifold from both of the unscreened phases. Terminations of excited states occur when the excited-state manifold on one side of the transition can no longer be reached from the ground state on the other side. This happens in Fig.\ \ref{fig:excitations_single_channel}b, where the high-$S_z$ unscreened excited state (blue line) can be excited from the partially screened ground state, but not from the unscreened low-$S_z$ ground state, reflecting the selection rule $\Delta Q=\pm 1$.

Corresponding results for $S_i=3/2$ are shown in Figs.\ \ref{fig:excitations_single_channel}c and d. Excitations out of the unscreened ground states are shown in panel c. For negative $D$, there is only a single excitation into the singly-screened states, while the anisotropy splitting appears as two excitations at positive $D$. This effectively leads to the termination of one of the excitations at the quantum phase transition. A similar effect is present for excitations out of the screened ground states shown in Fig.\ \ref{fig:excitations_single_channel}d. Here, the anisotropy splitting appears only for excitations out of the high-spin screened ground state (green background; left term diagram), but not for excitations out of the low-spin screened ground state (red background; right term diagram). In both cases, the reason for the termination is associated with the selection rules for tunneling excitations, albeit now with a selection rule for the spin projection $S_z$. Electron tunneling can change $S_z$ only by $\pm \frac{1}{2}$. The transitions into both anisotropy-split sublevels satisfy this selection rule for excitations out of the high-spin partially screened ground state (see left term diagram), but not for excitations out of the low-spin partially screened ground state (see right term diagram). Similarly, for transitions out of the unscreened states, the anisotropy splitting of the singly-screened states is only compatible with the $S_z$ selection rule, when exciting out of the low-$S_z$ ground state. 

A nonzero transverse anisotropy $E$ leads to qualitative changes in several of these results. First, a nonzero $E$ eliminates the quantum phase transition at $D=0$ in Fig.\ \ref{fig:excitations_single_channel}a. Second, it mixes states with $\Delta S_z=2$, splitting the degenerate unscreened $Q=0$ states with $S_z=\pm 1$ for $S_i=1$. When $Q=0$ is the ground-state manifold, this results in the appearance of an additional finite-temperature thermal peak. When $Q=0$ is the excited state, the YSR excitation splits into three anisotropy sublevels even at zero temperature. This would happen in Fig.\ \ref{fig:excitations_single_channel}b. No corresponding splitting appears in Fig.\ \ref{fig:excitations_single_channel}d for $S_i=3/2$, where the degeneracy of the $Q=0$ levels is protected by time reversal. Third, a nonzero $E$ leads to eigenstates, which are superpositions of different $S_z$ states. Consequently, there will generally be no forbidden transitions in Figs.\ \ref{fig:excitations_single_channel}c and d, and the anisotropy splitting  becomes visible for both signs of the anisotropy. 

\subsubsection{Parity-preserving excitations}
\label{sec:ppe}

Figure \ref{fig:ParityPreservingExcitations_SingleChannel} shows subgap spectra with parity-preserving excitations for different impurity spins $S_i$. Within our model, the parity-preserving spectra are independent of the superconducting gap as initial and final states involve the same number of paired channels. However, the pairing strength determines the range of energies, over which the excitations are observed as subgap excitations. We therefore plot the subgap spectra for both ground-state parities for the entire range of anisotropies. For a specific point in the phase diagram, the relevant excitation spectrum is the one for the particular ground-state parity and the particular $\Delta$ determines the energy range, over which the excitations appear at subgap energies.  

For the unscreened states (top row in Fig.\ \ref{fig:ParityPreservingExcitations_SingleChannel}), parity preserving excitations just reflect the anisotropy splitting of the impurity-spin spectrum. While the slopes of the states with the largest and smallest impurity-spin projections $S_z$ are symmetric on the two sides of the quantum phase transition at $D=0$, the slope of transitions into states with intermediate $S_z$ values differs between the two sides, reflecting the different ground states. 

For the singly screened states (bottom row in Fig.\ \ref{fig:ParityPreservingExcitations_SingleChannel}), the number of excitations remains unchanged compared to the screened state when the impurity spin is half-integer. In contrast, the number of excitations is reduced by one for integer impurity spins. These splittings are consistent with the expectation that screening results in an effective $(S_i-\frac{1}{2})$-spin. This means that the partially screened impurity behaves just like a quantum impurity with a correspondingly reduced spin, albeit with a renormalized anisotropy $D$ and noticeably nonlinear anisotropy dependence at larger positive $D$.  

The nonlinear dependence on the anisotropy $D$ is particularly pronounced for half-integer impurity spins and excitations into states with spin projections $S_z=1$. This can be understood as follows. At large and positive $D$, the spin projection of a half-integer impurity spin strongly prefers the $S_z=\pm \frac{1}{2}$ states. Then, the $S_z=0$ ground state becomes a linear combination of the states $|S_i,M=\pm\frac{1}{2}\rangle\otimes|\frac{1}{2},\mp\frac{1}{2}\rangle$, while the excited $S_z=\pm 1$ states are made up of $|S_i,M=\pm\frac{1}{2}\rangle\otimes|\frac{1}{2},\pm\frac{1}{2}\rangle$. All these states have the same anisotropy energy, and the corresponding excitation energy must saturate at large $D$, implying a nonlinear dependence on $D$. 

For partially screened phases, there are additional spin excitations at higher energies. For the parameters chosen here, such excitations appear for $S_i=1$ (dashed lines in Fig.\ \ref{fig:ParityPreservingExcitations_SingleChannel}). These emerge from the ferromagnetically coupled quasiparticle states with total spin $S=\frac{3}{2}$ at $D=0$. 
However, it should be noted that these are above-gap excitations. As can be seen in Fig.\ \ref{fig:phase_diagram_single_channel}, the maximal $\Delta$, at which there is a partially screened ground state for $J=1$ is $\Delta=1$, and these excitations appear only at higher energies. 

\subsection{Analytical considerations}

Many aspects of  the phase diagrams and the excitation spectra can be understood analytically by considering the limits of small and large uniaxial anisotropies. 

\subsubsection{Even-fermion-parity eigenstates}

The even-fermion-parity subspace is spanned by the unoccupied and doubly-occupied states of the single-site superconductor. Consistent with the spin-singlet nature of the superconductor, neither of these states has an effective spin, and the exchange coupling has no nonzero matrix elements within this subspace. Thus, the low-energy eigenstates are direct products of the impurity eigenstates $|S_i,M\rangle$ (with $M=-S_{i}, -S_{i}+1,\ldots, S_{i}-1, S_{i}$) and the paired electron state $|\mathrm{BCS}\rangle$, 
\begin{equation}
    |S_{i},S_z\rangle \otimes |\mathrm{BCS}\rangle,
\end{equation}
with eigenenergies 
\begin{equation}
    E_e(S_z) = - \Delta + DS_z^2.
\label{even_ferm-par_single_channel}
\end{equation}
Here, the first term is the pairing energy of the BCS ground state $|\mathrm{BCS}\rangle$, while the second term is the anisotropy energy (assuming zero transverse anisotropy $E=0$). Here and below, the subscript of the energy denotes the fermion parity and the argument indicates the projection of the total spin. We denote the projections of the total spin by $S_z$ and the projections of the impurity spin by $M$. These projections coincide for the even-fermion-parity subspace considered here, but are in general different.  

A nonzero transverse anisotropy $E$ will further split the degeneracy between the states with $\pm S_z$ for integer impurity spins. In contrast, this is a Kramers degeneracy protected by time reversal symmetry for half-integer spins.

\subsubsection{Odd-fermion-parity eigenstates at small uniaxial anisotropies}

Unlike for the even-fermion-parity subspace, the spectrum of the odd-fermion-parity subspace cannot be found analytically in full generality. However, much insight can be gained from perturbation theory in the limits of small and large anisotropies. We begin with the limit of small uniaxial anisotropy $D$. The odd-fermion-parity subspace is spanned by the states
\begin{equation}
   |S_{i}, M;\frac{1}{2},\sigma\rangle = |S_{i}, M\rangle \otimes |\frac{1}{2},\sigma\rangle.
\label{odd_basis}
\end{equation}
Initially neglecting the anisotropy, the rotationally symmetric exchange interaction couples impurity and quasiparticle spins into eigenstates of the total spin. For antiferromagnetic exchange interaction, the low-energy states have total spin $S_{i}-\frac{1}{2}$, leaving a $2S_{i}$-fold degenerate manifold of eigenstates. Using Clebsch-Gordan coefficients, these states can be expanded in the product basis in Eq.\ (\ref{odd_basis}), 
\begin{widetext}
\begin{eqnarray}
 |S_{i},\frac{1}{2};S_{i}-\frac{1}{2},S_z\rangle &=& \sum_{\sigma=\pm \frac{1}{2}} |S_{i}, S_z+\sigma;\frac{1}{2},-\sigma\rangle \langle S_{i}, S_z+\sigma ;\frac{1}{2},-\sigma|S_{i},\frac{1}{2};S_{i}-\frac{1}{2},S_z\rangle
 \nonumber\\
&=& \sqrt{\frac{1}{2}+\frac{S_z}{2S_{i}+1}} |S_{i}, S_z+\frac{1}{2};\frac{1}{2},-\frac{1}{2} \rangle-\sqrt{\frac{1}{2}-\frac{S_z}{2S_{i}+1}} |S_{i}, S_z-\frac{1}{2};\frac{1}{2},\frac{1}{2} \rangle.
\label{Jimpminonehalf}
\end{eqnarray}
\end{widetext}
Here, the first pair of entries in the state on the left hand side denote the spins that are being coupled and the second pair give the eigenvalues of the total spin and its projection. A standard calculation shows that these states have exchange energy $-\frac{J}{2}(S_{i}+1)$. 

This $2S_{i}$-fold degenerate manifold of eigenstates is split further by the uniaxial anisotropy $D$. Perturbatively in $D$, the states (\ref{Jimpminonehalf}) remain eigenstates of the Hamiltonian with energies
\begin{equation}
   E_o(S_z) = -\frac{J}{2}(S_{i}+1) +D \left(  \frac{2S_{i}+3}{2S_{i}+1} S_z^2  +\frac{1}{4}\right).
\label{odd_ferm-par_single_channel}
\end{equation}
Note that the anisotropy energy is nonzero even for vanishing projection of the total spin, $S_z=0$, a situation which can occur for half-integer impurity spins $S_i$. The reason is that according to Eq.\ (\ref{Jimpminonehalf}), the $S_z=0$ states are composed of states with $M=\pm \frac{1}{2}$. We will see below that this fact has ramifications for the phase diagrams.

\subsubsection{Odd-fermion-parity eigenstates at large uniaxial anisotropies}

In the opposite limit of large anisotropy, we first consider $D<0$. The low-energy eigenstates of the anisotropy Hamiltonian, $|S_{i}, \pm S_{i};\frac{1}{2},\sigma\rangle$, are shifted in energy by the exchange coupling. To linear order in $J$, only the longitudinal exchange coupling $JS_{i,z}s_z$ contributes and we find
\begin{equation}
    E_o(\pm (S_{i}-\frac{1}{2})) \simeq DS_{i}^2 -\frac{J S_{i}}{2}.
\label{odd_large_negative_D}
\end{equation}
Comparing to Eq.\ (\ref{odd_ferm-par_single_channel}), we observe that the large and negative anisotropy effectively reduces the gain in exchange energy. 

For large and positive uniaxial anisotropy $D$, we need to distinguish between integer and half-integer impurity spins. For integer impurity spins, the lowest-energy eigenstates of the uniaxial anisotropy are $|S_{i},M=0\rangle\otimes|\frac{1}{2},\sigma\rangle$ and have zero anisotropy energy. These two states are Kramers partners and will therefore remain uncoupled by the exchange interaction. However, the transverse exchange coupling shifts their energy in second order in perturbation theory, and we find
\begin{equation}
   E_o(\pm\frac{1}{2})\simeq -\frac{J^2 S_{i}(S_{i}+1)}{4D}.
\label{odd_largeD_integer}
\end{equation}
For large positive $D$, the exchange interaction is thus considerably less effective in binding quasiparticles to integer impurity spins. 

For half-integer impurity spins, there are four low-energy eigenstates $|S_{i}, M=\pm \frac{1}{2}\rangle \otimes |\frac{1}{2},\sigma\rangle$ of the uniaxial anisotropy, with anisotropy energy $D/4$. Unlike for integer impurity spins, time reversal symmetry no longer forbids exchange couplings within this degenerate manifold and the longitudinal exchange coupling contributes in addition to transverse exchange. We thus find a splitting in first order in the exchange coupling, and the state binding a quasiparticle has energy
\begin{equation}
    E_o(0) \simeq \frac{D}{4} - \frac{J}{2}(S_{i}+1).
\label{odd_largeD_halfinteger}
\end{equation}
At large and positive $D$, quasiparticles are therefore much more effectively bound to half-integer impurity spins than to integer impurity spins. This is not offset by the anisotropy energy as this energy also contributes to the energy of the unscreened state. 

\subsubsection{Phase diagrams}

We are now in a position to interpret the phase diagrams in Fig.\ \ref{fig:phase_diagram_single_channel}. Quantum phase transitions occur whenever two ground states cross in energy. On the one hand, this happens at $D=0$. The ground state favors large projections $S_z$ of the total spin at negative uniaxial anisotropy $D$, but small $S_z$ at positive $D$. This quantum phase transition occurs strictly at $D=0$. It is absent, when the total spin is smaller than $1$ and the anisotropy does not split the eigenstates, e.g., for a spin-$\frac{1}{2}$ impurity or a singly screened spin-1 impurity. It is present for vanishing transverse anisotropy $E$ only. 

On the other hand, quantum phase transitions are associated with the binding of a quasiparticle to the impurity. These occur when states with different fermion parities cross in energy. Thus, for the single-channel model the phase boundary is determined by the condition $E_e=E_o$. First consider $D<0$, where the anisotropy favors large projections of the impurity spin. For the unscreened state with even fermion parity, Eq.\ (\ref{even_ferm-par_single_channel}) gives
\begin{equation}
    E_e \simeq -\Delta +D S_{i}^2
\end{equation} 
for the lowest-energy state. For large and negative $D$, this should be compared to the lowest-energy odd-fermion-parity state, whose energy $E_o$ is given in Eq.\ (\ref{odd_large_negative_D}), and the quantum phase transition occurs at
\begin{equation}
     \Delta = \frac{JS_{i}}{2}.
\end{equation}   
For small and negative $D$, the screened state with odd fermion parity extends to larger $\Delta$. In this case, Eq.\ (\ref{odd_ferm-par_single_channel}) with $S_z=\pm(S_{i}-\frac{1}{2})$ gives
\begin{equation}
    E_o \simeq -\frac{J}{2}(S_{i}+1) - D\left(1-S_{i}^2 -\frac{2}{2S_{i}+1}\right)
\end{equation} 
for the lowest-energy state with odd fermion parity. Thus, the phase boundary follows 
\begin{equation}
     \Delta = \frac{J}{2}(S_{i}+1) +D \left(1-\frac{2}{2S_{i}+1}\right).
\label{phase_boundary_neg_Delta}
\end{equation}   
in this region.

For $D>0$, the anisotropy favors small values of the spin projection $S_z$, and there are qualitative differences between integer and half-integer impurity spins. At small uniaxial anisotropy $D$, Eq.\ (\ref{even_ferm-par_single_channel}) implies that the lowest-energy state with even fermion parity has energy
\begin{equation}
      E_e \simeq -\Delta +  \begin{dcases}  0  & \text{integer $S_{i}$;} \\
       \frac{D}{4} &  \text{half-integer $S_{i}$.}
\end{dcases}
\end{equation}
The difference emerges from the fact that the minimal spin projection is $S_z=0$ for integer impurity spins, but $S_z=\pm\frac{1}{2}$ for half-integer impurity spins. Similarly, we find from Eq.\ (\ref{odd_ferm-par_single_channel}) that the lowest-energy state with odd fermion parity has 
\begin{equation}
      E_o \simeq -\frac{J}{2}(S_{i}+1) +  \begin{dcases} \frac{S_{i}+1}{2S_{i}+1} D  & \text{integer $S_{i}$;} \\
           \frac{D}{4}  & \text{half-integer $S_{i}$.} \end{dcases} 
\end{equation}
Due to the screening electron, the minimal spin projection is now equal to $S_z=\pm \frac{1}{2}$ for integer and $S_z=0$ for half-integer impurity spins $S_{i}$. As a result of the fact that anisotropy contributes to the energy for half-integer $S_{i}$ despite the fact that $S_z=0$ [see discussion below Eq.\ (\ref{odd_ferm-par_single_channel})], the phase boundary remains unaffected by small anisotropies for half-integer spins,
\begin{equation}
   \Delta \simeq  \frac{J}{2}(S_{i}+1).
\end{equation}
In contrast, small anisotropies favor the unscreened state for integer impurity spins and the phase boundary follows
\begin{equation}
  \Delta \simeq \frac{J}{2}(S_{i}+1) - \frac{S_{i}+1}{2S_{i}+1} D
\end{equation}
Comparing to Eq.\ (\ref{phase_boundary_neg_Delta}), the slope of the phase boundary not only changes sign at $D=0$ resulting in a cusp, but also has different magnitudes for positive and negative $D$.

Finally, at large positive $D$, we find from Eqs.\ (\ref{even_ferm-par_single_channel}), (\ref{odd_largeD_integer}), and (\ref{odd_largeD_halfinteger}) that the phase boundary is given by 
\begin{equation}
   \Delta \simeq  \begin{dcases} \frac{J^2}{4D} S_{i}(S_{i}+1)  & \text{integer $S_{i}$;}\\
   \frac{J}{2} (S_{i}+1)  & \text{half-integer $S_{i}$.}
\end{dcases}
\end{equation}
For half-integer spins, the phase boundary occurs at the same value of $\Delta$ as for small positive anisotropy $D$. This is consistent with the numerical results in Fig.\ \ref{fig:phase_diagram_single_channel}, which show that for $D>0$, the phase boundary between unscreened and screened ground states of half-integer spin impurities is completely independent of $D$ (but see also the discussion in Sec.\ \ref{sec:beyond} below). In contrast, large positive anisotropy suppresses the screened ground state substantially for integer impurity spins, with the phase boundary approaching $\Delta=0$ for $D\to\infty$.  

\subsubsection{Excitation spectra}

Excitation energies follow directly from the eigenenergies computed above. Moreover, Fermi's Golden Rule allows one to compute dependences of tunneling rates for instance on the anisotropy. Here, we briefly discuss the selection rule $\Delta S_z=\pm\frac{1}{2}$ for tunneling excitations with $\Delta Q = 1$. There are two issues. First, in the case of metal coordination complexes, tunneling will typically proceed by cotunneling via a virtual state of the coordination complex, instead of direct tunneling into the substrate \cite{Anderson1966}. Second, tunneling into subgap states of superconductors can be a single-electron process followed by inelastic excitation, or a two-electron Andreev process transferring a Cooper pair into the superconducting substrate \cite{Ruby2015}. 

The selection rule $\Delta S_z=\pm\frac{1}{2}$ evidently applies to direct single-electron tunneling into (or out of) the superconducting substrate from a nonmagnetic STM tip. In this process, dominant for low tunneling rates, the tunneling electron excites the subgap YSR quasiparticle in the substrate, followed by rapid inelastic excitation of  the YSR quasiparticle into the above-gap continuum \cite{Ruby2015}. The tunneling Hamiltonian changes the spin projection due to the tunneling electron, but does not involve the impurity spin operator. Thus, a nonzero tunneling amplitude originates from those components of the inital and final states which have the same projections $M$ of the impurity spin. 

Cotunneling into the superconducting substrate via virtual states of the metal coordination complex generally proceeds via several interfering tunneling paths including both, potential and exchange scattering \cite{Appelbaum1967,Farinacci2020}. The effective tunneling Hamiltonian 
\begin{equation}
  H_T = \sum_{\sigma\sigma^\prime}\psi_\sigma^\dagger(\mathbf{R}) [ V_T \delta^{\phantom\dagger}_{\sigma\sigma^\prime} + J_T \mathbf{S}_i \cdot {\mathbf{s}}^{\phantom\dagger}_{\sigma \sigma^\prime}]\phi^{\phantom\dagger}_{\sigma^\prime}(\mathbf{R}) + \mathrm{h.c.}
\end{equation}
emerges from eliminating virtual high-energy states of the metal coordination complex by a Schrieffer-Wolff transformation. Here, we denote the electron operators in the tip and in the substrate at the tip position $\mathbf{R}$ by $\phi_{\sigma}(\mathbf{R})$ and $\psi_{\sigma}(\mathbf{R})$, respectively. Potential scattering $V_T$ leaves the impurity spin unchanged, just as direct tunneling into the superconducting substrate. In contrast, the amplitude for exchange scattering $J_T$ depends on the impurity spin operator. The selection rule $\Delta S_z=\pm \frac{1}{2}$ applies to both contributions, even though the impurity spin projection can be altered in the exchange process. For instance, for tunneling from tip to substrate, the $S_{i,+}s_-$ term adds a spin-down electron to the substrate, while changing the impurity spin projection by $+1$. 

The importance of the transverse contributions $S_{i,+}s_-+S_{i,-}s_+$ to the exchange scattering depends on the uniaxial anisotropy. Their contribution will be strongly suppressed in the limit of large negative anisotropy $D$ for any impurity spin and for large positive $D$ for integer impurity spins. 

At large tunneling rates, inelastic processes are no longer efficient at exciting YSR quasiparticles into the quasiparticle continuum. In this case, tunneling resonances associated with YSR states are predominantly due to resonant Andreev processes, which transfer Cooper pairs into the superconducting substrate \cite{Ruby2015}. The expression for the tunneling current due to resonant Andreev processes involves the Fermi golden rule rates for electron and hole processes \cite{Ruby2015,Acero2020} and thus the same matrix elements as the single-electron tunneling processes. As a result, the resonant Andreev processes are subject to the same selection rule for the total-spin projection as the single-electron processes. 

\subsection{Beyond the single-site approximation}
\label{sec:beyond}

Higher-spin impurities coupled to a single superconducting channel in the presence of uniaxial anisotropy have been investigated in extensive numerical renormalization group (NRG) calculations in Ref.\ \cite{Zitko2011}. The paper contains {\em schematic} phase diagrams as well as representative excitation spectra for the unscreened and partially screened phases, which can be directly compared with the results of our model. 
 
Our phase diagrams in Fig.\ \ref{fig:phase_diagram_single_channel} are in close correspondence with schematic phase diagrams for $S_i=1,\frac{3}{2}$, and $2$ obtained from the NRG calculation (Fig.\  2 of Ref.\ \cite{Zitko2011}). The NRG calculations also predict that uniaxial anisotropy of either sign suppresses the partially screened state for the integer impurity spins $S_i=1,2$. Moreover, the phase diagrams agree in that for the half-integer impurity spin $S_i=\frac{3}{2}$, uniaxial anisotropy suppresses the partially screened phase for $D<0$, but not for $D>0$. In the latter case, however, there is a difference. Figure\ \ref{fig:phase_diagram_single_channel} shows a phase boundary, which is independent of $D$, while NRG predicts that positive uniaxial anisotropy even favors the partially screened case. 

This difference can be traced to the neglect of Kondo renormalizations in our model. These renormalizations are themselves sensitive to the anisotropy $D$ and may thus affect the phase boundary. For half-integer spins, the Kondo renormalizations are stronger at large positive $D$ than at $D=0$, implying that the screened phases become increasingly favored as $D>0$ increases. The underlying reason is the twofold ground-state degeneracy of the uncoupled impurity between the states with $S_z=\pm\frac{1}{2}$ that remains at large $D>0$, which contrasts with the  $(2S_i+1)$-fold degeneracy at zero anisotropy, $D=0$. While the Kondo renormalizations from energies above the anisotropy scale are identical in both cases, there are differences on scales between the anisotropy scale and the superconducting gap.

These differences can be quantified by considering the scaling equations for the exchange coupling as a function of the band cutoff $\Lambda$ in the two limits. For vanishing $D$, the scaling equations take the form \cite{Anderson1970}
\begin{equation}
    \frac{dJ}{d\ln \Lambda} = -J^2
\end{equation}
and apply to the single-channel Kondo effect for any $S_i$. For large and positive $D$, the twofold degeneracy implies that the problem maps to a spin-$\frac{1}{2}$ model with anisotropic exchange interactions $J_z=J$ and $J_\perp = (S_i+\frac{1}{2})J$. The anisotropy arises because $S_{i,z}\ket{S_i,\pm\frac{1}{2}}=\pm \frac{1}{2}\ket{S_i,\pm\frac{1}{2}}$, but $S_{i,\pm}\ket{S_i,\mp\frac{1}{2}}=(S_i+ \frac{1}{2})\ket{S_i,\pm\frac{1}{2}}$. Thus, spin-flip processes become relatively more important at large $D>0$, implying a flow to larger exchange couplings. This can also be seen explicitly from the corresponding scaling equations \cite{Anderson1970}
\begin{eqnarray}
    \frac{dJ_z}{d\ln \Lambda} &=& -J_\perp^2 \nonumber\\
    \frac{dJ_\perp}{d\ln \Lambda} &=& -J_z J_\perp.
\end{eqnarray}
These predict a Kondo temperature
\begin{equation}
    T_K=\Lambda_0 e^{-\frac{2}{\sqrt{J_\perp^2-J_z^2}}\arctan{\sqrt{\frac{J_\perp-J_z}{J_\perp+J_z}}}},
\end{equation}
which increases with $J_\perp>J_z$ \cite{Zitko2008}. 

For integer spins and $D>0$, the Kondo renormalizations will effectively terminate once the cutoff becomes comparable to the anisotropy splitting. Thus, the renormalized exchange coupling decreases as $D$ increases, further amplifying the suppression of the screened phases by the anisotropy. 

The dependence of Kondo renormalizations on $D$ will also affect the phase boundaries at negative $D$ for any spin. As $D$ becomes large, the two states $\ket{S_i,\pm S_i}$ with the largest spin projections increasingly dominate. Spin-flip scattering between these states is only possible in higher-order perturbation theory, leading to a suppression of the Kondo renormalizations. This should further suppress the screened phases at large and negative $D$.

\begin{figure*}[]
\includegraphics[width=0.98\textwidth]{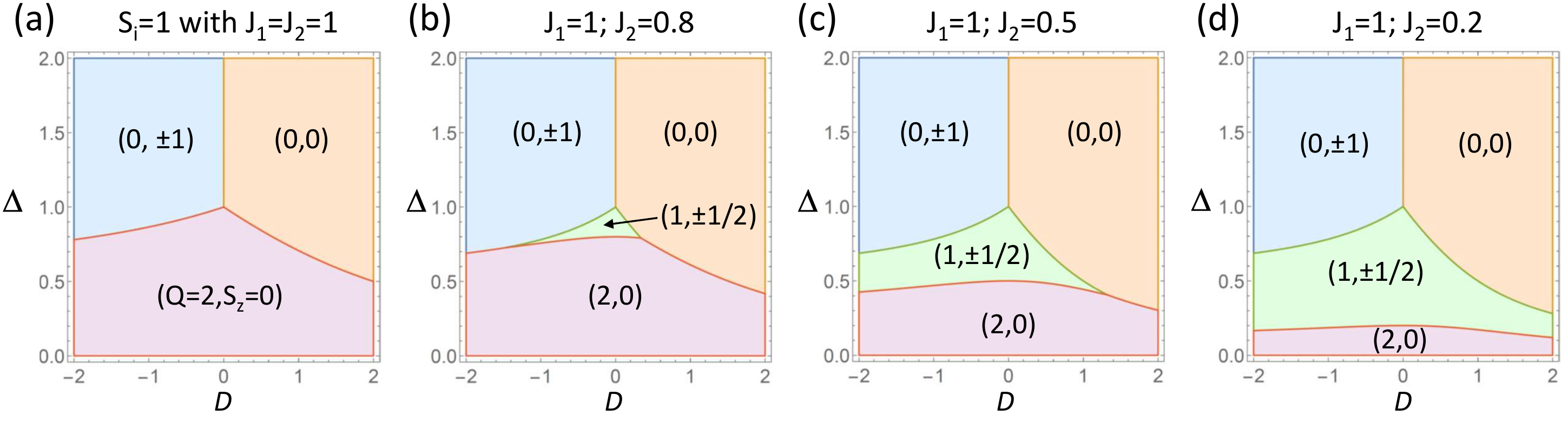}
	\caption{Phase diagrams for an impurity spin with $S_{i}=1$ coupled to two channels as a function of pairing strength $\Delta$ and uniaxial anisotropy $D$, for different values of the exchange couplings $J_1$ and $J_2$ in the two channels. Transverse anisotropy is set to zero. Phases are labeled by $(Q,S_z)$. (a) For equal couplings of both channels, there are direct quantum phase transitions between the unscreened phases $(0,1)$ and $(0,0)$ to the fully screened phase $(2,0)$. (b) For similar, but different exchange couplings in the two channels, a singly-screened phase $(1,\pm\frac{1}{2})$ appears at small anisotropy, while there is a direct transition between the unscreened and doubly screened phases at larger anisotropy of either sign. (The singly-screened phase reappears for intermediante $\Delta$ at even larger $D<0$, see text.) (c) For significantly different exchange couplings (right panel), the intermediate singly-screened phase exists for all negative $D$, but only up to a maximal positive $D$. (d) For very asymmetric exchange couplings, the singly screened phase persists for all  values of the anisotropy $D$.}
	\label{fig_S1_PhaseDiagrams}
\end{figure*}

There is also a remarkable degree of correspondence of the excitation spectra in Figs.\ \ref{fig:excitations_single_channel} and \ref{fig:ParityPreservingExcitations_SingleChannel} with the corresponding NRG results (Fig.\ 14 in Ref.\ \cite{Zitko2011}). In comparing these results, it should be noted that we separate excitations into tunneling spectra and parity-preserving excitations, and account for selection rules of the tunneling process. When accounting for these differences of presentation, agreement includes the appearance of zero-energy and finite-energy cusps of the excitation energies at quantum phase transitions, the quantum numbers of excitations, and in many cases even relative slopes of excitation energies as function of $D$. 

Based on this comparison, we conclude that most (though not all) aspects of the phase diagrams and the excitation spectra are controlled by the spin couplings, which are fully included in our model. In principle, it would be desirable to combine the approach taken here with a more complete treatment of Kondo renormalizations, but this is beyond the scope of the present paper. 

\section{YSR states of magnetic adatoms: Multi-channel model}

In view of the success of the model for single-channel situations, we now turn to the general case, in which all $2S_{i}$ channels are exchange coupled to the impurity spin. As argued above, this will typically happen for magnetic adatoms, with the site symmetry potentially enforcing equal exchange couplings for sets of symmetry-related channels. 

\subsection{Phase diagrams}

When all $2S_{i}$ channels are exchange coupled to the impurity spin, the impurity can in principle bind a quasiparticle in each of these channel. The impurity spin will be completely unscreened when binding no quasiparticle, fully screened when binding a quasiparticle in each of the $2S_{i}$ channels, or partially screened when the number $Q$ of bound quasiparticles is in between zero and $2S_{i}$. Since fermion parity is separately conserved in every channel, ground states with different numbers $Q$ of bound quasiparticles define separate phases of the system. In the absence of transverse anisotropy, we can further classify the ground states according to the projection $S_z$ of the total spin, so that we label phases by $(Q,S_z)$.

Figure \ref{fig_S1_PhaseDiagrams} shows phase diagrams for $S_i=1$ as a function of the pairing strength $\Delta$ and the uniaxial anisotropy $D$ (and vanishing transverse anisotropy, $E=0$) for representative values of  the exchange couplings in the two channels. For equal exchange couplings, $J_1=J_2=1$, the phase diagram exhibits direct transitions between phases with unscreened spin at large pairing energy $\Delta$ and a fully screened phase at smaller $\Delta$, see Fig.\ \ref{fig_S1_PhaseDiagrams}a. There are two unscreened phases $(0,1)$ and $(0,0)$, which differ in their total-spin projections, reflecting the different signs of the uniaxial anisotropy. In contrast, there is only one fully screened phase $(1,0)$. An intermediate singly-screened phase $(1,\pm\frac{1}{2})$ appears when the two channels involve different exchange couplings, and becomes increasingly prominent as the ratio between the exchange couplings increases, see Figs.\ \ref{fig_S1_PhaseDiagrams}b-d. Positive uniaxial anisotropy $D$ suppresses the screened phases more strongly than a negative $D$ of the same magnitude. In particular, positive anisotropy can completely  suppress the singly-screened phase for sufficiently large $D>0$, see Figs.\ \ref{fig_S1_PhaseDiagrams}b-c. For negative anisotropies, the singly-screened phase can disappear at intermediate values, but will always reappear for sufficiently large and negative $D$ (not shown; see analytical considerations below). 

\begin{figure}[t]
	\includegraphics[width=0.98\columnwidth]{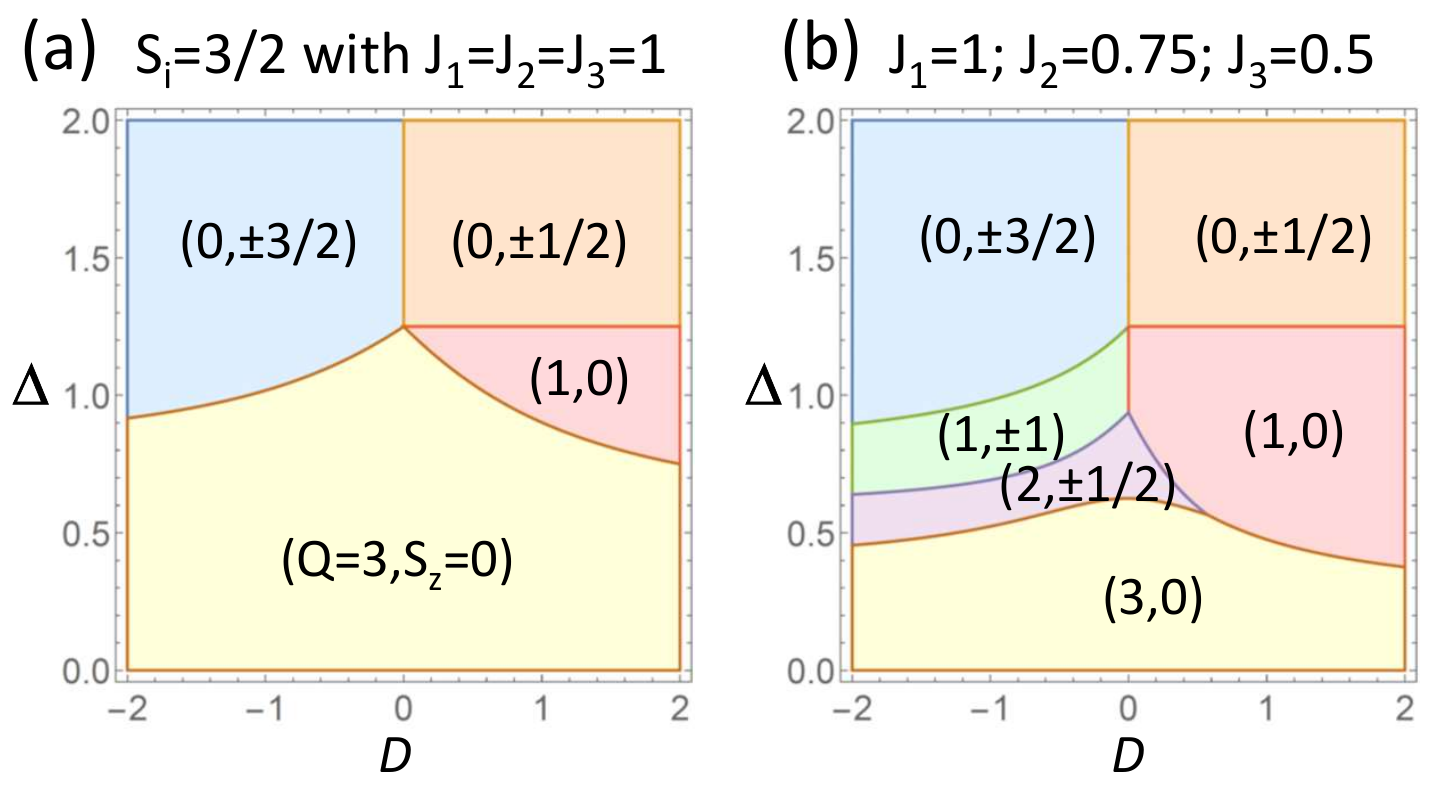}
	\caption{Phase diagrams for $S_{i}=\frac{3}{2}$ coupled to three channels as a function of pairing strength $\Delta$ and uniaxial anisotropy $D$. Transverse anisotropy is set to zero. Phases are labeled by $(Q,S_z)$. (a) When the exchange couplings in all three channels are equal, there is a direct transition from the unscreened to the fully screened state for negative uniaxial anisotropy, $D<0$. For positive anisotropy, $D>0$, there is an intermediate phase in which the impurity is screened only in a single channel. (b) If the exhange couplings differ between all channels, there is a cascade of phases at negative anisotropy $D$, in which the impurity spin is successively screened as the pairing strength is reduced. The phases correspond to states in which $Q=0,1,2$, or $3$ quasiparticles are bound to the impurity. In contrast, for sufficiently large positive anisotropy $D$, there are only phases with zero, one, or three bound quasiparticles.}
	\label{fig:S32_PhaseDiagrams}
\end{figure}

\begin{figure*}[]
	\includegraphics[width=0.98\textwidth]{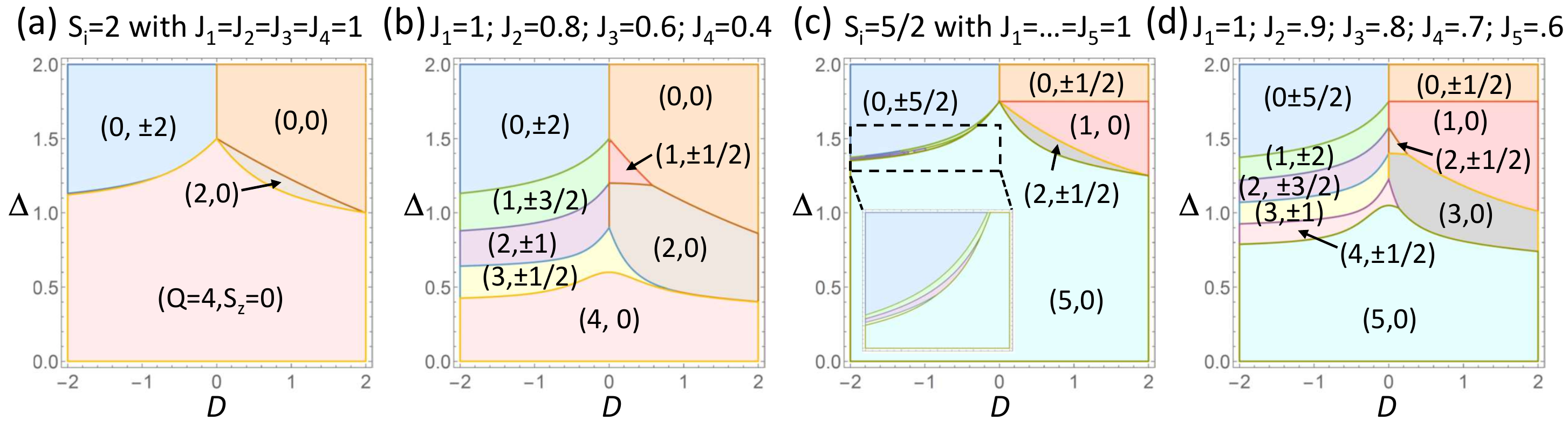}
	\caption{Phase diagrams for higher impurity spins with (a,b) $S_{i}=2$ and (c,d) $S_{i}=\frac{5}{2}$ as a function of pairing strength $\Delta$ and uniaxial anisotropy $D$. The values of the exchange couplings in all $2S_i+1$ channels are given above the panels. Transverse anisotropy is set to zero. Phases are labeled by $(Q,S_z)$. The inset in panel c shows a higher-resolution calculation of the $D<0$ region between the unscreened and the fully screened phases. The color coding of the intermediate phases is as in panel d.}
	\label{fig:HighSpin_PhaseDiagrams}
\end{figure*}

Representative phase diagrams for a half-integer impurity spin, namely $S_i=\frac{3}{2}$, are shown in Fig.\ \ref{fig:S32_PhaseDiagrams}. First consider the case of equal exchange couplings in all three channels, shown in Fig.\ \ref{fig:S32_PhaseDiagrams}a. At negative uniaxial anisotropies $D$, there is a direct transition between the unscreened phase $(0,\pm \frac{3}{2})$ at larger $\Delta$ and a fully screened phase $(3,0)$ at smaller $\Delta$. In contrast, the unscreened and screened phases are separated by an intermediate singly-screened phase $(1,0)$ at positive anisotropies $D>0$. As for the single-channel case discussed above, the phase boundary between the unscreened and singly-screened phase is independent of $D$ for positive anisotropies $D>0$. When the exchange couplings differ between the channels (see Fig.\ \ref{fig:S32_PhaseDiagrams}b), there is a cascade of phases as a function of $\Delta$ for negative $D$. These phases bind $Q=1$, $2$, or $3$ quasiparticles to the impurity and have a spin projection, which is reduced by $\frac{1}{2}$ by each bound quasiparticle. In contrast, there is still a direct transition between the singly- and triply screened phases at a sufficiently large and positive $D$. The doubly-screened phase is found to persist to arbitrarily large positive $D$, when reducing the smallest exchange coupling $J_3$ below a certain threshold value (not shown). 

Phase diagrams for yet larger impurity spins are shown in Fig.\ \ref{fig:HighSpin_PhaseDiagrams}. These phase diagrams for $S_i=2$ (panels a and b) and  $S_i=\frac{5}{2}$ (panels c and d) show similar characteristic features as the corresponding phase diagrams for smaller impurity spins. When all exchange couplings are equal, the phase diagram is dominated by the unscreened and fully screened phases for negative $D$. However, there appears a very narrow region between these two phases, in which phases with other $Q$ appear. These are not resolved in the main panels, but can be seen in the inset of Fig.\ \ref{fig:HighSpin_PhaseDiagrams}c. When the exchange couplings in all channels differ, there is a cascade of transitions through phases $(Q,\pm(S_i-\frac{Q}{2}))$, in which the impurity spin is increasingly screened by quasiparticles. At positive $D$, the phase diagram is systematically dominated by phases with zero projection of the total spin, $S_z=0$. With the exception of the fully unscreened phase, phases with half-integer impurity spin appear only in much smaller regions and disappear for sufficiently large and positive $D$. 

\subsection{Excitation spectra}

\subsubsection{Tunneling spectra} 

\begin{figure*}[]
	\includegraphics[width=0.98\textwidth]{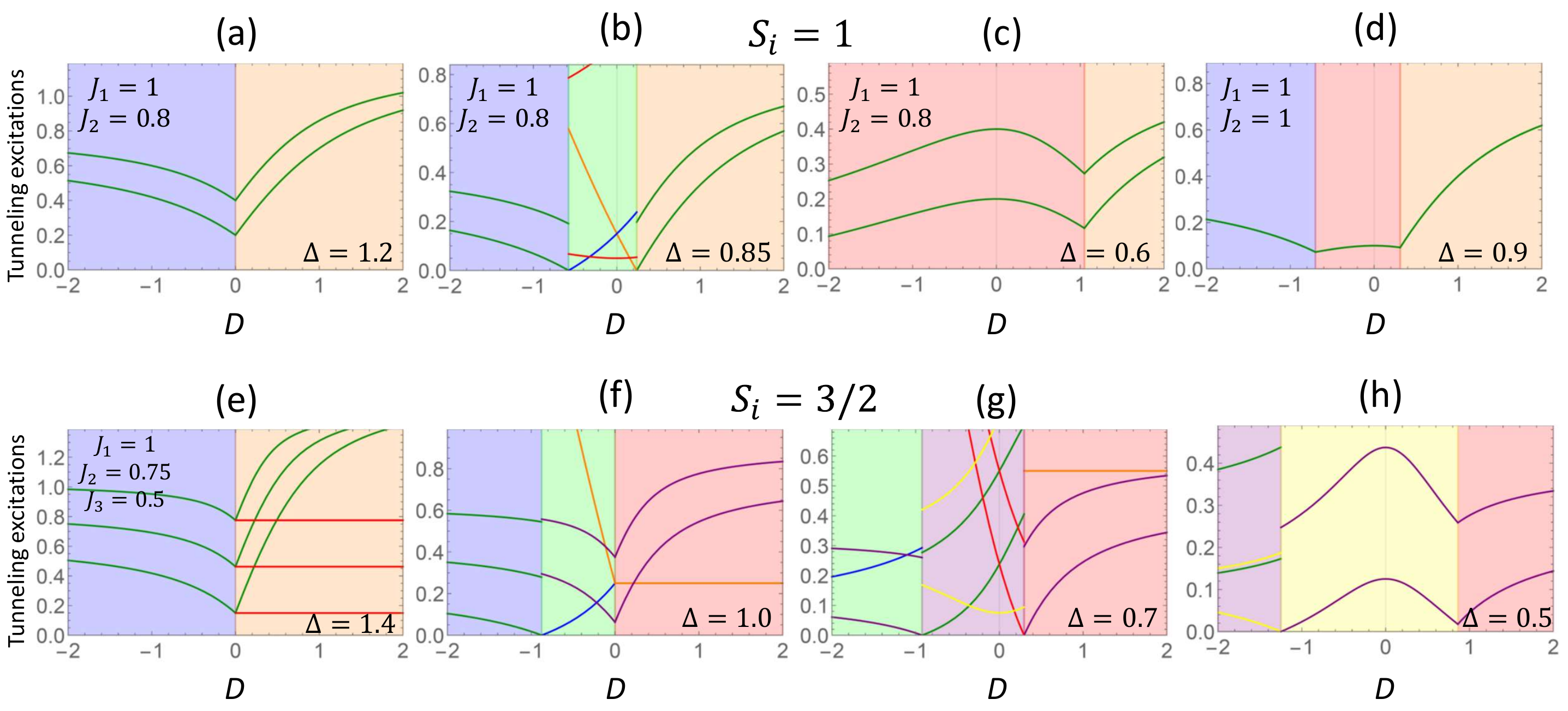}
	\caption{Subgap tunneling spectra of the multi-channel model for (a-d) $S_{i}=1$ and (e-h) $S_{i}=\frac{3}{2}$ as a function of uniaxial anisotropy $D$, for pairing strengths $\Delta$ and exchange couplings $J_i$ as indicated in the panels. Transverse anisotropy is set to zero. Background color indicates the ground state and line color the excited state (same color coding as in the corresponding phase diagrams in Figs.\ \ref{fig_S1_PhaseDiagrams}) and (\ref{fig:S32_PhaseDiagrams}). The maximal excitation energy shown corresponds to the gap edge $\Delta$ of the superconductor.
}
	\label{Fig_ExcitationSpectra_MultiChannel}
\end{figure*}

Figure \ref{Fig_ExcitationSpectra_MultiChannel} shows subgap tunneling spectra for $S_i=1$ and $S_i=\frac{3}{2}$ for representative values of the pairing strength $\Delta$ and the exchange couplings $J_i$. As for the single-channel case, tunneling excitations occur between states with $\Delta Q=\pm 1$ and $\Delta S_z=\pm \frac{1}{2}$. First consider the excitation spectra for $S_i=1$  (top row of Fig.\ \ref{Fig_ExcitationSpectra_MultiChannel}). For unequal exchange couplings of the two channels, there are two tunneling excitations out of the unscreened ground states, one for each channel (Fig.\ \ref{Fig_ExcitationSpectra_MultiChannel}a). The quantum phase transition at $D=0$ is reflected in finite-energy cusps in both excitations. The excitation energies of the two channels become degenerate for equal couplings to the two channels  (Fig.\ \ref{Fig_ExcitationSpectra_MultiChannel}d). 

At intermediate pairing strength $\Delta$, a singly screened ground state appears (green background in Fig.\ \ref{Fig_ExcitationSpectra_MultiChannel}b). In this region, tunneling can excite the unscreened states (blue and orange lines) or the doubly screened state (red lines). The excitations into the unscreened states exhibit an anisotropy splitting as for the single-channel case and a zero-energy cusp at the quantum phase transitions, at which the excited state turns into the ground state. One of the excitations into the doubly screened state appears at rather low energies due to the close proximity of the doubly screened phase at this particular value of $\Delta$ (cp.\ Fig.\ \ref{fig_S1_PhaseDiagrams}b). 
Interestingly, there are two excitations into the doubly screened phase at $D<0$. 

This is a direct consequence of the quantum nature of the impurity spin and can be best understood from the limit of large and negative $D$. In this limit, there are two fully screened states, $|S_i=1,M=1\rangle\otimes|\frac{1}{2},-\frac{1}{2}\rangle\otimes|\frac{1}{2},-\frac{1}{2}\rangle$ and $|S_i=1,M=-1\rangle\otimes|\frac{1}{2},\frac{1}{2}\rangle\otimes|\frac{1}{2},\frac{1}{2}\rangle$. The two states are almost degenerate, but will be coupled at order $J_1J_2/D$ in second-order perturbation theory in the transverse-exchange couplings. This leads to a splitting and the two red lines in Fig.\ \ref{Fig_ExcitationSpectra_MultiChannel}b extend from these weakly split levels at large and negative $D$. 

At smaller pairing strength $\Delta$, there is a direct phase transition between the doubly-screened and an unscreened phase (Fig.\ \ref{Fig_ExcitationSpectra_MultiChannel}c). These phases differ by $\Delta Q=2$ and are therefore not associated with a zero-energy cusp of a tunneling excitation. Instead, the tunneling transitions, one for each channel, excite into the singly-screened states from both phases, which results in finite-energy cusps in the excitation spectra.  

Many of these themes also appear in the subgap tunneling spectra for $S_i=\frac{3}{2}$ shown in Fig.\ \ref{Fig_ExcitationSpectra_MultiChannel}e-h. Excitations from the unscreened into the singly-screened states can now proceed within three channels, so that the tunneling spectrum in Fig.\ \ref{Fig_ExcitationSpectra_MultiChannel}e is a threefold copy of the one for the single-channel model shown in Fig.\ \ref{fig:excitations_single_channel}c. Tunneling excitations from the singly screened ground states (green and red background in Fig.\ \ref{Fig_ExcitationSpectra_MultiChannel}f) can now also proceed into the doubly screened states (purple lines). Since channel 1 is already screened in the ground state, there are two such excitations, one each for channels 2 and 3. 

Close to the quantum phase transition between the high-spin singly screened and unscreened phases (green and blue backgrounds in Fig.\ \ref{Fig_ExcitationSpectra_MultiChannel}f, respectively), the excitations from the singly screened into doubly screened states (purple lines) are rather close in energy to the excitations from the unscreened into the singly screened states in channels 2 and 3 (green lines). This reflects that both sets of lines appear due to exciting a quasiparticle in channels 2 and 3. However, in the singly screened phase, the impurity spin is already partially screened by a quasiparticle in channel 1, and the additional quasiparticle in channel 2 or 3 interacts with this partially screened impurity. For a purely classical impurity, interacting only via the longitudinal exchange interaction with the quasiparticle, this would be inconsequential and the excitation spectrum would be continuous across the quantum phase transition. In contrast, for a quantum impurity, the transverse exchange interactions modify this picture and lead to a nonzero shift of the excitation energies at the quantum phase transition. 

At even smaller pairing strength (Fig.\ \ref{Fig_ExcitationSpectra_MultiChannel}g), the doubly screened ground state appears around zero anisotropy (purple background). Tunneling excitations from this state into the singly screened states (green and red lines) exhibit anisotropy splittings. In addition, there are tunneling excitations into the fully paired state (yellow lines). At first sight, there should only be a single such excitation since channel 3 is the only unpaired channel in the doubly-screened ground state. However, we observe a pair of such excitations for negative $D$, which rapidly move closer in energy as $D$ becomes large and negative. The origin of this doubling can again be understood from the limit of large negative $D$. In this limit, there are two fully screened states, $|S_i=\frac{3}{2},M=\frac{3}{2}\rangle\otimes|\frac{1}{2},-\frac{1}{2}\rangle\otimes|\frac{1}{2},-\frac{1}{2}\rangle\otimes|\frac{1}{2},-\frac{1}{2}\rangle$ and $|S_i=\frac{3}{2},M=-\frac{3}{2}\rangle\otimes|\frac{1}{2},\frac{1}{2}\rangle\otimes|\frac{1}{2},\frac{1}{2}\rangle\otimes|\frac{1}{2},\frac{1}{2}\rangle$. These states are again coupled at order $J_1J_2J_3/D^2$ in the transverse exchange couplings, leading to a splitting. The two yellow lines extend again from these weakly split states in the limit of large and negative $D$. The same splitting can also be seen in Fig.\ \ref{Fig_ExcitationSpectra_MultiChannel}h.

\subsubsection{Parity-preserving excitations} 

\begin{figure*}[t]
	\includegraphics[width=1.\textwidth]{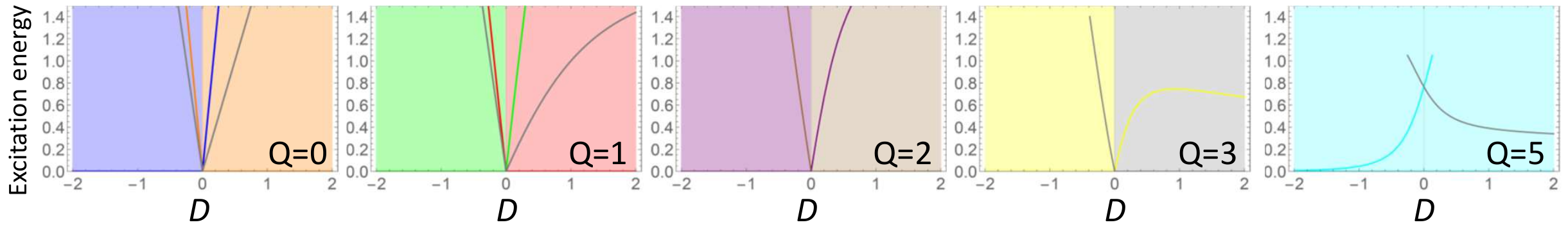}
	\caption{Subgap excitation spectra of the multi-channel model at fixed fermion parity for $S_{i}=\frac{5}{2}$ as a function of anisotropy $D$. The exchange couplings were chosen as  $J_1=1$, $J_2=0.9$, $J_3=0.8$, $J_4=0.7$, and $J_5=0.6$. Transverse anisotropy is set to zero. Background color indicates the ground state and line color the excited state (same color coding as in the corresponding phase diagram in Fig.\ \ref{fig:HighSpin_PhaseDiagrams}d), except for the gray lines for $Q=0,1,5$. The gray lines in the panels for $Q=0$ and $1$ correspond to intermediate total spin projections $S_z=\frac{3}{2}$ and $S_z=1$, respectively, which do not become the ground state at any $D$. The gray line in the panel for $Q=5$ corresponds to an $S_z=1$ excitation. There is no panel for $Q=4$, since there are no subgap excitations in that case. The excitation spectra are independent of pairing strength $\Delta$, except that $\Delta$ determines the range, over which the state still appears as a subgap excitation. The excitations are limited to energies below the maximal $\Delta$, at which the particular ground state can still be observed at $D=0$.}
	\label{Fig_ParityPreservingExcitations_MultiChannel}
\end{figure*}

Figure \ref{Fig_ParityPreservingExcitations_MultiChannel} shows the subgap excitations at fixed fermion parity for $S_i=\frac{5}{2}$. For $Q$ bound quasiparticles, the excitation spectrum can be understood as the spectrum of the anisotropy Hamiltonian of a spin-$(S_i-\frac{Q}{2})$ impurity, consistent with the antiferromagnetic coupling between the impurity spin and the conduction electrons. The nonlinear dependence on uniaxial anisotropy can be understood as explained in Sec.\ \ref{sec:ppe} above for the single-channel model. 

Additional low-energy excitations appear for $Q=5$. As already discussed above for $S_i=1$ and $S_i=\frac{3}{2}$, there are two low-energy states $|S_i=\frac{5}{2},M=\frac{5}{2}\rangle\otimes|\frac{1}{2},-\frac{1}{2}\rangle^{\otimes 5}$ and $|S_i=\frac{5}{2},M=-\frac{5}{2}\rangle\otimes|\frac{1}{2},\frac{1}{2}\rangle^{\otimes 5}$ at large negative $D$. These two $S_z=0$ states are coupled in fifth-order perturbation theory in the transverse exchange couplings, resulting in a splitting of order $J_1J_2J_3J_4J_5/D^4$. The excitation from the ground to the excited state of this pair leads to the (cyan) excitation, which rapidly approaches zero energy as $D$ becomes large and negative. At $D=0$, the ground state evolves into the total-spin eigenstate $|S=0,S_z=0\rangle$, while the excited state evolves into $|S=1,S_z=0\rangle$. The latter is part of a degenerate triplet, together with $|S=1,S_z=\pm 1\rangle$. It is these latter states, from which the two-fold degenerate gray line develops for $Q=5$.

\subsection{Analytical considerations}

Many aspects  of the multi-channel model can again be readily understood analytically by considering the limits of large (negative and positive) or zero anisotropies. We begin with large and negative anisotropies, which favor large projections of the impurity spin, $S_{i,z}=\pm S_i$. Then, the many-body energy of a state with $Q$ screened channels is given by
\begin{equation}
    E_B  = -\Delta(2S_{i}-Q) +DS_{i}^2 - \frac{1}{2}(J_1+\ldots+J_Q)S_i.
\end{equation}
Here, the first term accounts for the superconducting pairing energy in the $2S_{i}-Q$ unpaired channels, the second term is the anisotropy energy of the impurity spin, and the last term denotes the exchange coupling between the impurity and quasiparticle spins in the $Q$ screened channels with exchange couplings $J_1,\ldots,J_Q$. Notice that the transverse terms in the exchange coupling take the impurity spin out of the low-energy manifold $S_{{i},z}=\pm S_{i}$, so that only the longitudinal contribution to the exchange coupling contributes to leading order in the limit of large and negative $D$. Equating these many-body energies for states with $Q$ and $Q+1$ bound quasiparticles, we find that an additional quasiparticle is bound by the impurity whenever 
\begin{equation}
   \Delta = \frac{J_m}{2}S_{i}.
\label{phaseboundarylargenegativeD}
\end{equation}
This implies a direct transition from the unscreened into the fully screened phase, when all exchange couplings $J_m$ take on the same value and a cascade of transitions binding an additional quasiparticle each as $\Delta$ decreases, when all $J_m$ are different. This is in agreement with the phase diagrams in Figs.\ \ref{fig_S1_PhaseDiagrams}, \ref{fig:S32_PhaseDiagrams}, and \ref{fig:HighSpin_PhaseDiagrams}. The only apparent exception is the phase diagram for $S_{i}=1$ in Fig.\ \ref{fig_S1_PhaseDiagrams}b. However, as already alluded to above, the intermediate singly-screened phase reemerges at negative anisotropy $D$ of even larger magnitudes, consistent with the perturbative result in Eq.\ (\ref{phaseboundarylargenegativeD}). 

At large and positive $D$, the anisotropy favors minimal projections $S_{{i},z}$ of the impurity spin. For integer impurity spins, large and positive anisotropy $D$ locks the impurity into the state with $M=0$, and the exchange coupling with the quasiparticles only contributes in quadratic order as for the single-channel case discussed in Sec.\ \ref{sec:mol_phase_diagrams} above. Importantly, this implies that large and positive uniaxial anisotropy $D$ strongly favors the unscreened phase, with the critical  exchange couplings for quantum phase transitions to screened phases scaling as $J\sim \sqrt{\Delta D}$.  

For a single bound quasiparticle, the second-order processes are analogous to those for the single-channel case in Eq.\ (\ref{odd_largeD_integer}). For multiple bound quasiparticles, second-order processes in the transverse exchange coupling also induce simultaneous flips of antiparallel quasiparticle spins, in addition to diagonal energy shifts. The lowest-energy configurations will thus have minimal projections $S_z$ of the total spin. We illustrate the principle for the case of two bound quasiparticles. Minimal $S_z$ implies that the lowest-energy state lies in the subspace spanned by the quasiparticle states $|\frac{1}{2}\rangle \otimes |-\frac{1}{2}\rangle$ and $|-\frac{1}{2}\rangle \otimes |\frac{1}{2}\rangle$. Projected into this basis, the second-order terms in the exchange couplings take the form
\begin{equation}
     -\frac{S_{i}(S_{i}+1)}{4D}\left( \begin{array} {cc}  J_1^2+J_2^2 & 2J_1J_2 \\
        2J_1J_2  & J_1^2+J_2^2 \end{array} \right).
\label{eq_singlyscreenedintegerlargeDpositiveHamiltonian}
\end{equation} 
Here, the factor $2$ in the offdiagonal terms emerges from the two possible orders, in which the second-order spin flip can proceed. For exchange couplings $J_1\geq J_2$, we thus find the ground-state energy 
\begin{equation}
   E_{B=0} = -2\Delta
\end{equation}
for the unscreened phase, 
\begin{equation}
   E_{B=1} \simeq -\Delta - \frac{J_1^2}{4D}S_{i}(S_{i}+1)
\label{eq_singlyscreenedintegerlargeDpositive}
\end{equation}
for the singly-screened phase, and 
\begin{equation}
   E_{B=2} \simeq  - \frac{(J_1+J_2)^2}{8D}S_{i}(S_{i}+1)
\end{equation}
for the doubly-screened phase. A transition from the unscreened state into the singly-screened state is thus predicted at $\Delta \simeq \frac{J_1^2}{4D}S_{i}(S_{i}+1)$, and into the doubly-screened phase at $\Delta \simeq \frac{(J_1+J_2)^2}{8D}S_{i}(S_{i}+1)$. This implies that at large $D>0$, the transition into the singly-screened phase is preempted by a direct transition into the doubly-screened phase as long as $J_1/J_2\leq 1+\sqrt{2}$, in agreement with the phase diagrams in Fig.\ \ref{fig_S1_PhaseDiagrams}.  It is also interesting to note that the ground state of the doubly-screened phase couples the two quasiparticle spins into an $M=0$ triplet state, which then forms a product state with the $M=0$ state of the impurity spin. These considerations can be readily extended to binding more than two quasiparticles to understand the corresponding regions in the phase diagrams for $S_{i}=2$ in Fig.\ \ref{fig:HighSpin_PhaseDiagrams}.

For half-integer impurity spins, large positive $D$ leads to a two-dimensional low-energy manifold of the impurity spin, spanned by $|M=\pm\frac{1}{2}\rangle$ and having anisotropy energy $D/4$. In this case, both the longitudinal and transverse exchange couplings contribute already in linear order and screened phases persist to much larger pairing energies compared to the integer-spin case. It is in principle straight-forward to perform perturbative calculations, but the resulting expressions are not particularly revealing.

In the limit of weak anisotropy, exchange coupling $Q$ quasiparticles to the impurity results in the eigenstates $|S_{i}-\frac{Q}{2},M\rangle$ (with $2S_{i}-Q$ unscreened channels). These eigenstates can be expanded into product states 
\begin{widetext}
\begin{equation}
   |S_{i},M;\sigma_1,\ldots,\sigma_Q\rangle = 
|S_{i},M\rangle \otimes |\frac{1}{2},\sigma_1\rangle \otimes \cdots \otimes |\frac{1}{2},\sigma_Q\rangle \otimes |\mathrm{BCS}\rangle^{\otimes  2S_{i}-Q}.
\end{equation}
Here, we write the product state on the right hand side such that the $Q$ screened channels follow the impurity state, and the unscreened channels are written last. Notice also that on the left hand side, we only write the spin projections of the quasiparticles for simplicity. Coupling the quasiparticles one by one, we find
\begin{eqnarray}
  |S_{i}-\frac{Q}{2},M\rangle = \sum_{\sigma_1,\ldots,\sigma_Q}  |S_{i},M+\sigma_1+\ldots+\sigma_Q;-\sigma_1,\ldots,-\sigma_Q\rangle   \langle S_{i},M+\sigma_1+\ldots+\sigma_Q;-\sigma_1,\ldots,-\sigma_Q | S_{i}-\frac{Q}{2},M\rangle ,\,\,\,\,\,\,\,\,\,\, 
  \end{eqnarray}
where the overlaps on the right hand side can be written as products of Clebsch-Gordan coefficients,
\begin{eqnarray}
  && \langle S_{i},M+\sigma_1+\ldots+\sigma_Q;-\sigma_1,\ldots,-\sigma_Q | S_{i}-\frac{Q}{2},M\rangle = 
       \langle S_{i},M+\sigma_1+\ldots+\sigma_Q;\frac{1}{2},-\sigma_1 |S_{i},\frac{1}{2}; S_{i}-\frac{1}{2},M +\sigma_2+\ldots+\sigma_Q\rangle \nonumber\\
  && \,\,\,\,\,\,\,\,\, \times
\langle S_{i}-\frac{1}{2},M+\sigma_2+\ldots+\sigma_Q;\frac{1}{2},-\sigma_2 |S_{i}-\frac{1}{2},\frac{1}{2}; S_{i}-1,M +\sigma_3+\ldots+\sigma_Q\rangle   \nonumber\\
  && \,\,\,\,\,\,\,\,\, \times \cdots \times
\langle S_{i}-\frac{Q-1}{2},M+\sigma_Q;\frac{1}{2},-\sigma_Q |S_{i}-\frac{Q-1}{2},\frac{1}{2}; S_{i}-\frac{Q}{2},M \rangle.
  \end{eqnarray}
\end{widetext}
In the absence of single-ion anisotropy, the many-body energy of these states is equal to 
\begin{equation}
    E_B(M)  = -\Delta(2S_{i}-Q) - \frac{1}{2}(J_1+\ldots+J_Q)(S_i+1).
\end{equation}
Thus, we find that the quantum phase transition, at which a particular channel $m$ is screened, occurs when
\begin{equation}
   \Delta = \frac{J_m}{2}(S_i+1).
\label{PhaseBoundaryZeroAniso}
\end{equation}
Consequently, there is also a cascade of transitions at zero anisotropy, binding quasiparticles one by one unless some channels have the same exchange couplings. We also conclude by comparison with Eq.\ (\ref{phaseboundarylargenegativeD}) that at large and negative uniaxial anisotropy, the corresponding quantum phase transitions occur at a smaller value of $\Delta$, i.e., large and negative anisotropy suppresses the screened phases. 

\section{Conclusion}

In experiment, YSR excitations in real metals come in a bewildering variety, which contrasts with the simple behavior of popular theoretical models. The number of YSR excitations varies widely between different magnetic adsorbates, or even for identical adsorbate and substrate, but differing surface orientations or adsorption sites. The most popular model treats the magnetic impurity as classical, while experiments show clear quantum behavior. Moreover, the model ignores the transition-metal nature of the impurity with its higher spin, the ligand- and crystal fields, and the magnetic anisotropy. The present paper is an attempt to transcend the standard modeling, in an effort to clarify the origins of the rich phenomenology of YSR excitations, while retaining a sufficiently simple framework to admit detailed understanding and to incorporate many experimentally relevant situations. 

The impurity spin is a result of the large Hund coupling within the $d$ shell of the magnetic ion, and may further depend on the ratio between Hund coupling and ligand- or crystal fields. The latter split the $d$ orbitals in accordance with the point group of the magnetic adsorbate. Each half-filled $d$ orbital is exchange coupled to its own symmetry-adapted conduction-electron channel, with exchange constants of equal magnitudes for symmetry-related $d$ orbitals. The impurity spin is in general subject to magnetic anisotropy, ultimately resulting from spin-orbit coupling. 

There can be a subgap YSR quasiparticle bound in any of the $2S_i$ conduction-electron channels. States with and without bound quasiparticles are distinguished by the fermion parity quantum numbers of the channels, which are all separately conserved. Tunneling excitations change the fermion parity in one of the channels. In the absence of anisotropy, one thus finds a maximal number of $2S_i$ YSR excitations, five, for instance, for the maximal possible spin $S_i=\frac{5}{2}$ of a transition-metal ion. When point-group symmetry imposes equal exchange couplings, some of these excitations will be degenerate, resulting in fewer than $2S_i$ separate YSR peaks in a tunneling experiment. 

Magnetic anisotropy leads to refinements of this picture. Assuming full spin rotation symmetry, the states in the absence of magnetic anisotropy can be further classified by their total spin and its projection. For antiferromagnetic exchange couplings between the impurity spin and the conduction electrons, binding $Q$ quasiparticles to the impurity results in a total spin of $S=S_i-\frac{Q}{2}$, corresponding to (partial) screening of the impurity spin. These $(2S_i-Q)$-fold degenerate spin multiplets split under the magnetic anisotropy.   
Kramers theorem imposes a twofold degeneracy of all states when $S_i-\frac{Q}{2}$ is half integer. Degeneracies can also occur for integer $S_i-\frac{Q}{2}$, when transverse anisotropy is symmetry forbidden. While the splitting patterns can be qualitatively understood in a picture, in which one subjects a spin-$(S_i-\frac{Q}{2})$ impurity to magnetic anisotropy, a quantitative understanding must account for the fact that the magnetic anisotropy acts only on the impurity spin. 

Magnetic anisotropy can lead to additional subgap YSR excitations in tunneling spectra. 
In the absence of anisotropy, the excitation count does not depend on whether the exchange coupling binds quasiparticles in any of the channels or not. In contrast, the number of excitations observed in the presence of anisotropy can depend sensitively on the particular ground and excited states as well as the magnitude of the anisotropy splitting. First, anisotropy splittings will only be observed at subgap energies as long as the splitting remains sufficiently small compared to the superconducting gap. Second, at zero temperature, the initial state of a tunneling process is always the lowest-energy state within the anisotropy-split ground-state manifold, but the final state can in principle be any of the sublevels of the anisotropy-split excited-state manifold. Since states with and without bound quasiparticles split into different numbers of sublevels, the detailed excitation count will now depend on the specific fermion-parity nature of the ground state. At finite temperatures, states other than the lowest-energy sublevel are also thermally populated, adding additional excitations with activated temperature dependence. 

The detailed excitation count is further governed by additional selection rules. In particular, the system remains symmetric under spin rotations about the $\hat z$-axis for uniaxial magnetic anisotropy. In this case, states can be classified according to the projection $S_z$ of the total spin. Tunneling excitations are then subject to the selection rule $\Delta S_z=\pm \frac{1}{2}$. This selection rule applies regardless of whether the YSR line is excited in a single-electron tunneling process or in a two-electron Andreev process. It also applies regardless of whether the tunneling amplitude is independent of the impurity spin as for direct tunneling into the substrate or dependent on the impurity spin as in cotunneling via virtual states associated with the impurity ion. The intensities with which particular excitations are being observed may, however, be quite sensitive to the particular tunneling process. For instance, while direct tunneling into the substrate leaves the impurity spin projection unchanged, this is not the case for the impurity-spin-dependent cotunneling process. 
 
The detailed phase diagram depends on the exchange couplings in the various channels. When all exchange couplings are substantially different, there will be ground states that bind any number $Q$ of quasiparticles, depending on pairing strength and anisotropy. Partially screened states are considerably more robust for negative (easy-axis) uniaxial anisotropy. For positive (easy-plane) uniaxial anisotropy, there are characteristic differences between half-integer and integer impurity spins. (Partially) screened phases are more robust for impurities with half-integer spin, but even in this case there is a stronger tendency towards skipping even-$Q$ phases than at $D<0$. In contrast, positive (easy-plane) uniaxial anisotropy, which is large compared to the exchange coupling between impurity spin and conduction electrons, substantially weakens the binding of quasiparticles to integer impurities, strongly suppressing all (partially) screened phases. When symmetry imposes that the exchange couplings of several channels are identical, these channels will tend to be screened simultaneously, or near simultaneously. The quantum phases of the model are not only characterized by $Q$, but also by the projection of the total spin, when transverse anisotropy is symmetry forbidden. The associated quantum phase transitions between phases with equal $Q$ occur at $D=0$, with a phase favoring maximal spin projections for easy-axis anisotropy, $D<0$, and a phase favoring minimal spin projections for easy-plane anisotropy, $D>0$. 

The quantum phase transitions between states with different $Q$ or different $S_z$ are reflected in the tunneling excitation spectra. In general, quantum phase transitions are signaled by zero-energy or finite-energy cusps in excitation energies as well as terminations of excitation lines. Zero-energy cusps occur for transitions between phases with neighboring values of $Q$, but are absent for transitions with $\Delta Q \neq 1$. Finite-energy cusps can exist in both cases. Excitations can terminate at a quantum phase transition, when the transition into a particular excited state is allowed on one side of the transition, but forbidden by selection rules on the other.

There are several aspects of this broad picture, which explicitly reflect the quantum nature of the impurity spin. One consequence is the suppression of the (partially) screened phases with increasing negative (easy-axis) uniaxial anisotropy $D$. At large and negative $D$, the impurity spin is effectively locked along the $\hat z$ direction and only the longitudinal exchange coupling contributes to the binding of quasiparticles. In contrast, longitudinal and transverse exchange couplings contribute for small $D$, resulting in more robust binding. Similarly, the binding of quasiparticles is entirely due to the transverse exchange coupling for positive (easy-plane) anisotropy. Another prominent consequence of the quantum nature of the impurity spin are the anisotropy splittings. There are also more subtle aspects. First, YSR excitations reflecting the binding or unbinding of a quasiparticle in some channel can exhibit discontinuities in excitation energy across a quantum phase transition. Second, the fully screened states at large negative $D$ exhibit a weak splitting induced by the transverse exchange coupling, which results in splittings of tunneling excitations and additional parity-preserving excitations. 

At the same time, our model neglects Kondo renormalizations. The broad qualitative agreement between our results and existing NRG calculations for the single-channel case implies that the structure of the phase diagrams as well as the subgap excitations are largely governed by the physics of spin couplings. However, we also observe differences. These can be readily understood in terms of the familiar scaling equations for the exchange couplings. Kondo renormalizations will cause the effective exchange couplings to grow with increasing easy-plane anisotropy $D>0$ for half-integer spins, but to decrease for integer spins. In contrast, the Kondo renormalizations are increasingly suppressed for any spin, as $D$ becomes large and negative. 
 
It is interesting to place existing experimental results within this broad picture. Magnetic adatoms frequently exhibit several YSR states \cite{Ji2008, Ruby2016, Choi2017, Liebhaber2019, Odobesko2020}, reflecting the coupling of the impurity spin to multiple conduction-electron channels. Direct evidence for the influence of crystal-field splittings on YSR excitations comes from spatially mapping the YSR wave functions of Mn adatoms on Pb surfaces by STM \cite{Ruby2016}. The maps as well as the number of YSR excitations are consistent with symmetry considerations for the particular surface orientations and adsorption sites. 

Experimentally exploring the phase diagrams and the quantum phase transitions is desirable but difficult for adatoms. YSR energies have been modified by varying the local density of states at the Fermi level exploiting quantum-well states \cite{Song2020}, a charge density wave \cite{Liebhaber2019}, or local surface reconstructions \cite{Odobesko2020}. In some cases, these variations were sufficiently large to induce a quantum phase transition \cite{Liebhaber2019}. However, these influences cannot be controlled in a continuous manner, but instead depend on the local properties of the adatom's adsorption site. In contrast, the distance of the STM tip to the substrate can be tuned continuously. The tip-induced electrostatic potential as well as mechanical forces may be used to manipulate the exchange coupling strength. This strategy has been successfully applied to molecular adsorbates (see below) \cite{Farinacci2018, Malavolti2018} and (sub)surface defects \cite{Huang2020, Chatzopoulos2021}.

On substrates with negligible spin-orbit coupling, anisotropy splittings would be signaled by identical spatial maps of different YSR excitations. However, there seem to be no reports of anisotropy splittings of YSR excitations for adatoms, presumably because there is only a narrow parameter window over which anisotropy splittings are small enough to yield additional subgap excitations and large enough to be experimentally resolvable. Under favorable conditions, the anisotropy parameters can be extracted from independent normal-state measurements of spin excitations as a function of applied magnetic field. Anisotropies on the order of the superconducting gap were extracted in this way for transition-metal impurities on Re substrates, but the associated inelastic spin excitations were located outside the superconducting gap \cite{Schneider2019}. 

While multiple YSR states are frequently observed for adatoms, transition-metal coordination compounds typically induce a single YSR excitation, if any. A YSR state present for one adsorption site on a particular substrate can be absent (or unresolved) for a different adsorption site of the same substrate \cite{Farinacci2020, Verdu2021}. This may be a consequence of a change in hybridization and thus exchange coupling between metal center and substrate or a change in sign or magnitude of the magnetic anisotropy. These changes may also be amplified by their effects on Kondo renormalizations. Multiple YSR states of metal coordination complexes have been interpreted in terms of a two-channel model in Ref.\ \cite{Verdu2021}, but are otherwise traced to anisotropy splittings \cite{Hatter2015}. Exploiting a moir\'e structure, Ref.\ \cite{Hatter2015} could realize both, unscreened  and screened ground states of a Mn-phthalocyanine molecule on Pb(111), which allowed for detection of the characteristic differences in the anisotropy splittings of the two ground states and the associated spectral weights. Unlike for adatoms, equal spatial STM maps do not constitute a signature of anisotropy-split YSR excitations for molecular adsorbates. For molecular adsorbates, tunneling does not map the YSR wave functions, but proceeds via virtual molecular orbitals and reflects their spatial dependence \cite{Farinacci2020}.  

\begin{figure*}[t]
	\includegraphics[width=.78\textwidth]{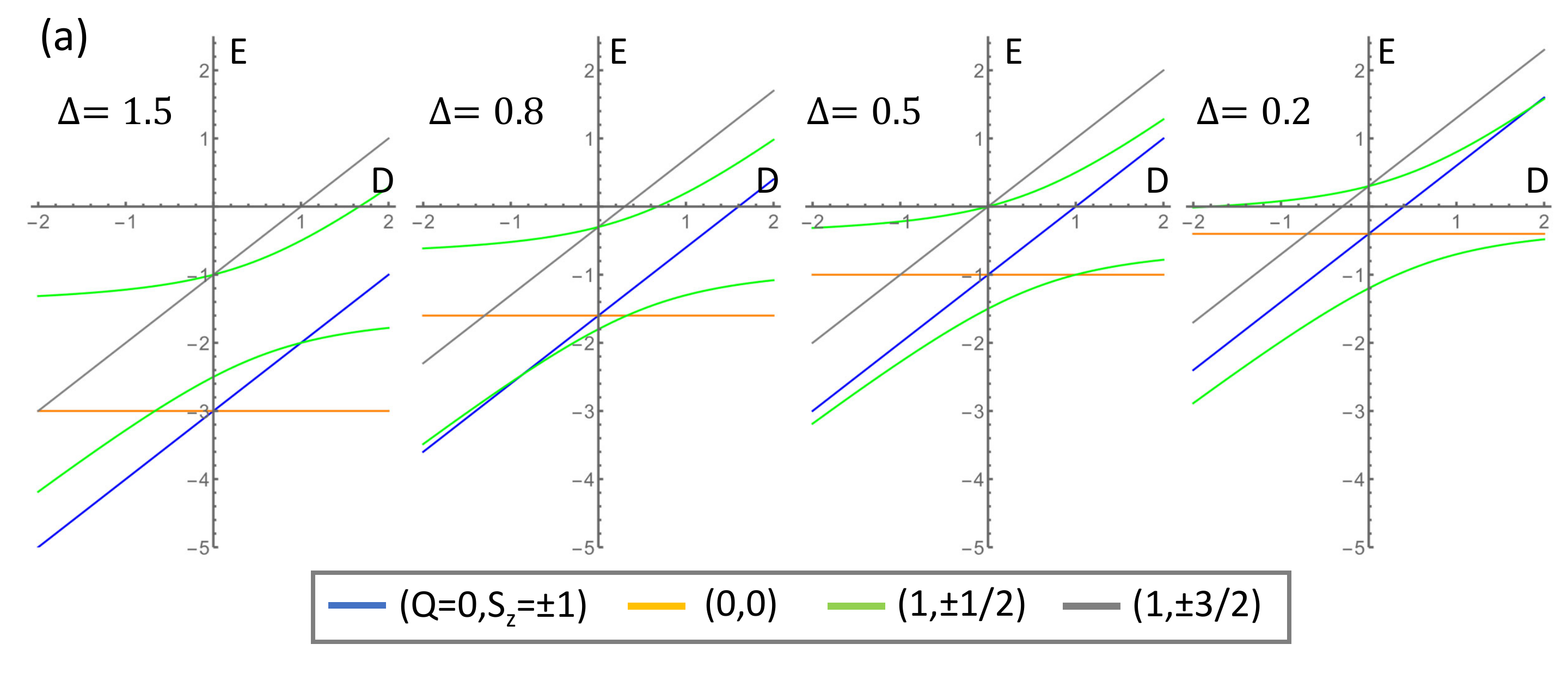}
	\includegraphics[width=.78\textwidth]{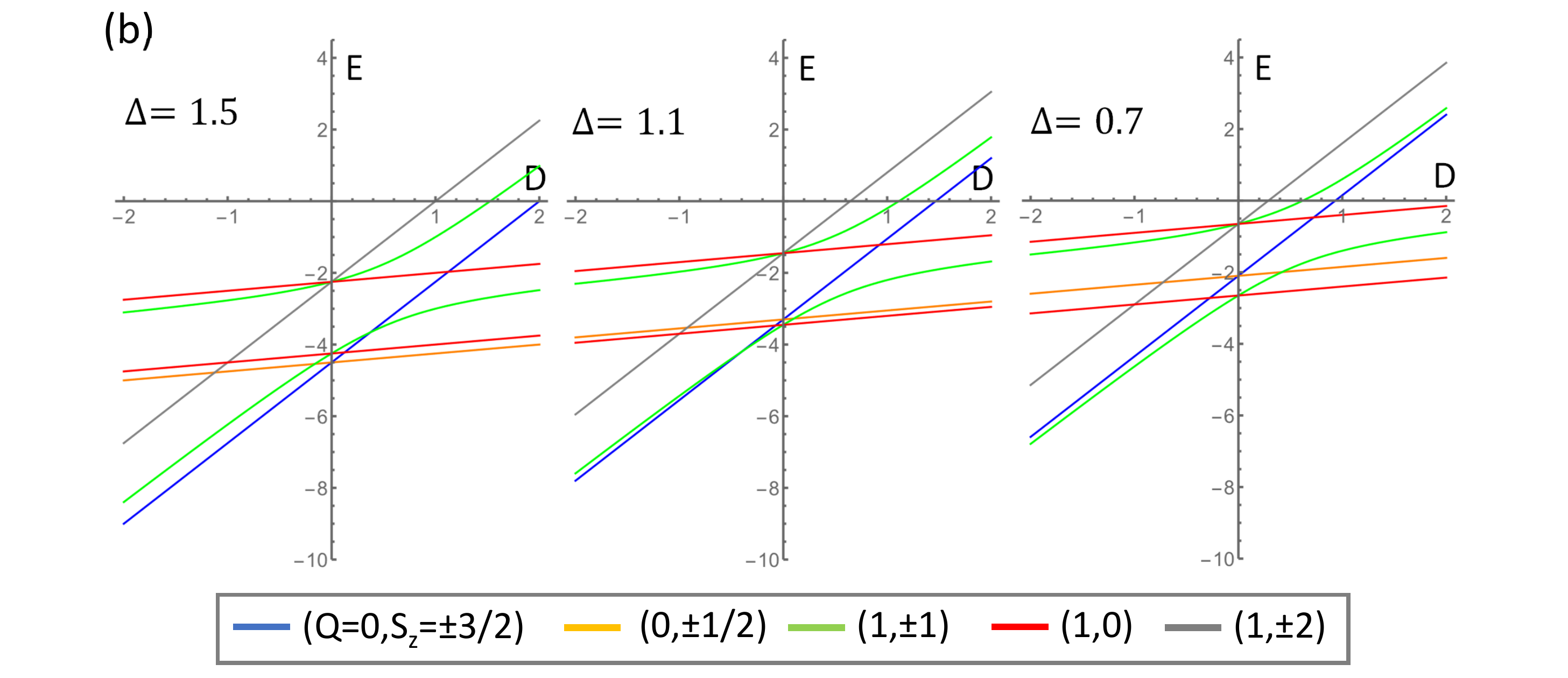}
	\caption{Representative plots of the eigenenergies $E$ of the single-channel Hamiltonian in Eq.\ (\ref{single_channel_hamiltonian}) for (a) $S=1$ and (b) $S=\frac{3}{2}$ as a function of anisotropy $D$ for $J=1$ and representative pairing strengths $\Delta$. (Transverse anisotropy $E=0$.) For even-fermion parity, spectra include only the BCS state  of the single-site superconductor. Colors label the number of bound quasiparticles $Q$ and the projection $S_z$ of total spin, see legends. Color coding matches the corresponding phase diagrams in Fig.\ \ref{fig:phase_diagram_single_channel}.  }
	\label{Fig_App}
\end{figure*}

Few experiments report the simultaneous observation of YSR states and above-gap spin excitations in the same system \cite{Kezilebieke2019,Schneider2019}. Our model does not preclude this in general. However, no spin excitations are expected when screening by binding quasiparticles results in an effective spin-$\frac{1}{2}$ or even spin-$0$ impurity. This situation should be particularly relevant for adatoms, which can be screened in multiple channels. Moreover, above-gap spin excitations exist, when the anisotropy is larger than the superconducting gap. However,  large positive (easy-plane) anisotropies substantially reduce the binding energy of YSR states for integer-spin impurities, potentially merging them with the BCS peak in experiment. It should, of course, be kept in mind that our model neglects the above-gap continuum, which plays a relevant role in the context of above-gap spin excitations. 

Existing experiments probing the quantum phase transitions were performed using metallic coordination complexes \cite{Hatter2015,Hatter2017,Farinacci2018, Malavolti2018}, and are thus limited to the case $\Delta Q=1$. The transitions could be induced by exploiting moir\'e lattices between superconducting substrate and adsorbed molecular structure \cite{Franke2011,Hatter2015,Hatter2017} or by explicitly manipulating the molecule-substrate interaction by exerting forces with the STM tip \cite{Farinacci2018,Malavolti2018}. It was identified by the observation of zero-energy cusps in the excitation spectrum, with  additional evidence provided by the discontinuous behavior of the peak strengths across the transition \cite{Hatter2015,Farinacci2018}. We also note that within experimental resolution, these experiments are consistent with true zero-energy cusps. Discontinuous changes in the YSR energies due to self-consistency effects have been predicted \cite{Salkola1997,Flatte1997,Meng2015}, but should be far below current experimental resolution for typical experimental parameters \cite{Farinacci2018}. There seem to be no existing experiments, which reveal quantum phase transitions with $\Delta Q\neq 1$. 

This work provides a starting point for further explorations of YSR excitations in real metals. Aspects of real metals that we have not considered include the existence of several bands \cite{Moca2008} or the effects of spin-orbit coupled substrates \cite{Noat2015,Sticlet2019,Glodzik2020,Liebhaber2019}. We have also sidestepped unconventional types of pairings \cite{Balatsky2006, Fan2020,Wang2021,Zhang2020}. These aspects need to be included, for instance, for a thorough understanding of YSR excitations of adsorbates and sub-surface impurities on superconducting transition-metal dichalcogenides such as NbSe$_2$, which have recently received significant attention \cite{Menard2015,Senkpiel2018,Liebhaber2019,Yang2020}. Another natural extension considers the couplings of multiple magnetic adsorbates \cite{Ruby2018, Choi2018, Kezilebieke2018,Kamlapure2018,Kamlapure2019,Beck2020}. 

\begin{acknowledgments}
We gratefully acknowledge discussions with Jacob Steiner and Christophe Mora as well as
funding by Deutsche Forschungsgemeinschaft through CRC 183 (project C03). 
\end{acknowledgments}

\appendix*

\section{Eigenvalue spectra} 

This appendix shows representative plots of the eigenenergies as a function of anisotropy $D$ for representative pairing strengths $\Delta$, see Fig.\ \ref{Fig_App}a and b for $S=1$ and $S=\frac{3}{2}$ impurity spins coupled to a single channel. The (symmetry-protected) level crossings of the lowest-energy states indicate quantum phase transitions between different ground states. Lower-energy states depend more sensitively on the anisotropy $D$ for easy-axis anisotropy (negative $D$ favoring large $S_z$) than for easy-plane anisotropy (positive $D$ favoring small $S_z$). Spectra focus on potential subgap states and include only the BCS state for even-fermion parity of the single-site superconductor. 

\bibliographystyle{apsrev4-1}

\end{document}